# Rapid Adaptive Matched Filter for Detecting Radar Targets with Unknown Velocity*

Anatolii A. Kononov, *Independent Researcher*

**Abstract**– This paper introduces a Doppler domain localized (DDL) implementation of the adaptive matched filter (AMF) for radar target detection in severely heterogeneous clutter environments with limited training data. The proposed detector uses the concept of a region of possible target detection (RPTD), a small set of Doppler bins that capture most of the target signal power. This RPTD-based DDL-AMF detector outperforms an earlier suggested DDL implementation of the generalized likelihood ratio (GLR) test, which employs the region of detection improvement (RODI) concept. Unlike the RODI-based DDL-GLR detector, the proposed DDL-AMF detector requires no information on clutter spectrum parameters and no measurements to determine the number and locations of RODIs. Moreover, the RODI-based DDL-GLR detector's performance falls far below the optimum when the target Doppler frequency is unknown. In contrast, the RPTD-based DDL-AMF detector ensures rapid adaptive detection with near-optimum performance under unknown target Doppler frequency and multimodal clutter spectra.

**Keywords**: Adaptive detection, AMF detector, Constant False Alarm Rate (CFAR), Doppler domain localized (DDL) adaptivity, GLR test, heterogeneous clutter, limited training data, rapid near-optimum detector

## I. INTRODUCTION

This paper addresses the challenge of achieving near-optimum detection performance for radar targets with unknown velocity in nonstationary and nonhomogeneous clutter environments with complex spectra. This is a fundamental problem for airborne and coastal surveillance radars, which must detect targets of interest amidst multiple strong interferences that vary in time and space. Reliable target detection in such scenarios is difficult due to the limited availability of training data for estimating unknown clutter statistics.

For simplicity, we focus on adaptive detectors for a monostatic pulse Doppler radar having a single antenna with no adaptive pattern control. We assume a coherent processing interval (CPI) with $N$ pulses, where $N$ must be sufficiently large even for the optimum detector (Neyman-Pearson's likelihood ratio test) to deliver a reliable detection of weak targets embedded in the strong clutter with complex spectra. In Gaussian interference, the classical adaptive detection algorithms, e.g., GLR test [1] and AMF [2], possess the desired CFAR property and may deliver near-optimum detection performance if they use at least as large $N$ as the optimum detector. Besides, they require a training data set containing at least $4N$ independent and identically distributed vectors [3]. However, even for moderate $N$, such training data are often unavailable in nonstationary and non-homogeneous clutter environments.

---

* This text is the original version of the manuscript authored by the researcher who obtained all the results presented here. Clarifying true authorship is to show respect for scientific ethics. Later, the author used this manuscript to publish the paper in *IEEE Access*, vol. 12, pp. 25411-25428, 2024, doi: 10.1109/ACCESS.2024.3363242.
In this text, Appendix E provides errata to the mentioned paper, aiming to ensure clarity and uphold the integrity of the published work for the research community.

Thus, directly implementing classical adaptive CFAR detectors does not solve the problem under consideration. To address this challenge, [4] proposed a Doppler domain localized generalized likelihood ratio (DDL-GLR) detector that combines the localized adaptivity principle in the Doppler domain with the GLR test. Implementing the DDL adaptive processing requires the Fourier transform (weighted or unweighted) must be first applied to the original time-domain data sampled at the output of the receiver I/Q channel to confine the signal and interference power to small regions in the Doppler domain.

According to [4], the key idea for implementing data-efficient adaptive detectors is to apply the adaptivity only to those areas of the Doppler domain where some detection improvement is necessary and possible. It proposed dividing the Doppler domain into two types of regions to identify such areas. The first type refers to the flat or nearly flat spectrum regions, and the second one refers to the sharply changing spectrum regions. Suppose the target Doppler frequency is within the first type of spectrum regions and far away from the clutter spectrum peaks. Then, the conventional processor employing a windowed DFT bank of filters followed by the traditional CA CFAR detectors is expected to provide near-optimum detection performance. In contrast, when the target Doppler falls on sharply changing spectrum regions or regions near the clutter spectrum peaks, the conventional detector performance may fall far below the optimum. Therefore, [4] suggested improving detection performance for such regions using adaptive processing. These regions are called the regions of detection improvement (RODI). Each RODI covers a small set of $n$ adjacent Doppler bins, with $n \ll N$.

As shown in [4], a RODI-based DDL-GLR detector may achieve near-optimum detection for $N \gg 1$ even when the RODI order is $4-6$. This small order results in an essentially smaller training data size than that for the time-domain GLR detector of order $N$. The DDL-GLR detector also has a lower computational load than its time-domain counterpart and allows parallel/distributed implementation. However, this detector has several flaws preventing its practical applications.

First, as pointed out in [4], a challenging problem is determining the number and locations of the RODIs to achieve near-optimum detection performance. Auxiliary clutter measurements are required to obtain the RODI parameters from the received data making the implementation of the RODI-based DDL-GLR detector complicated and computationally intensive. Additionally, this detector has the following disadvantages:

i) Determining the number and locations of the RODIs is overly complex when multiple clutter types

(e.g., ground, sea, and weather) appear in the cell under test. ii) Significant departures from optimum detection performance may occur even for known target Doppler frequencies, regardless of whether they take on or off DFT grid values. iii) These departures prevent the implementation of the RODI-based DDL-GLR detector under unknown Doppler frequencies by using test statistics associated with the regular DFT grid, as is done by a traditional CFAR algorithm when detecting targets embedded in thermal noise.

The present paper aims to develop a new adaptive DDL detector that is free of the disadvantages of the RODI-based DDL-GLR detector. The new adaptive DDL detector must ensure rapid and near-optimum detection of targets in training data-deficient scenarios without using RODIs, without information on clutter spectrum parameters, and without measurements for determining the number and locations of RODIs. It must also ensure reliable target detection in clutter environments with multimodal power spectral density (PSD), such as when multiple clutter types (ground, sea, rain, etc.) are simultaneously present in the cell under test. Additionally, it must provide near-optimum detection under unknown Doppler frequencies of targets of interest for any target frequency within the interval of unambiguous Doppler measurements.

This paper proposes a new DDL implementation of the classical time-domain AMF (TD-AMF) detector. This implementation employs the concept of a region of possible target detection (RPTD), a small set of Doppler bins that captures most of the target signal power. The proposed RPTD-based DDL-AMF detector is free of the disadvantages of an earlier suggested RODI-based DDL-GLR detector. The former ensures rapid adaptive detection with near-optimum performance under unknown target Doppler frequency and multimodal clutter spectra.

It should be noted that since the publication of [4], we have found only one work [5] on a similar topic in open literature. In addition, [6] extended the principle of adaptive DDL processing to the adaptive spatial-temporal (angle-Doppler) processing. One of the reasons for the lack of publications on adaptive DDL detection is that from the early 90s, much attention has been paid to the problem of adaptive target detection in sea clutter governed by compound-Gaussian (CG) distributions; [7-20] and references therein confirm the existence of plenty of publications on this topic.

The present paper considers the DDL-based adaptive radar detection under Gaussian interference for simplicity of discussion. Extending the theory of classical adaptive detection to the Doppler domain is straightforward since Gaussian distributions are closed under affine transformations. Building the theory of adaptive target detection in the Doppler domain under interference governed by a complex elliptically symmetric distribution (CES) is also simple: CES distribution is closed under affine transformations [21]. It should also be noted that the CES distributions class includes CG distributions as a particular case.

The remaining part of this paper is organized as follows. Section II describes the data representation for adaptive DDL detectors. Section III analyzes the disadvantages of the RODI-based DDL-GLR detector. Section IV introduces the concept of the region of possible target detection (RPTD). The new RPTD-based DDL-GLR and DDL-AMF detectors are then described in Section V. These are the detectors under investigation in the present paper. Section VI analyzes the detection performances of the detectors under investigation in several clutter scenarios under known and unknown target Doppler and discusses their advantages and disadvantages. Section VII evaluates the computational load of the RPTD-based DDL-AMF detector relative to its classical counterpart. The conclusion is given in Section VIII.

## II. Data Representation for Adaptive DDL Detectors

This section describes the data representation for adaptive DDL processing. To clarify implementing adaptive DDL detectors, it considers a data arrangement for two-dimensional pulse Doppler radar as a simple example.

Fig. 1 (a) illustrates a typical data arrangement for adaptive target detection in the time domain. The input data are organized into a complex-valued range-Pulse (fast/slow time) data matrix $\mathbf{X} = [x_{ij}]$ of size $M{\times}N$; $M$ is the number of range cells, and $N$ is the number of pulses in a current CPI. The entries in this matrix represent baseband demodulated, sampled, digitized, and pulse-compressed radar returns received in the CPI. In Fig. 1 (a), white circles mark the cell under test (CUT) samples that represent a CUT vector $\mathbf{x} = [x_{i1}\ x_{i2}\ \dots x_{iN}]^{\mathrm{T}}$ corresponding to the $i$-th range cell to be tested for the presence of targets. The shaded circles represent the samples in the training vectors $\mathbf{x}_q = \left[x_{i_q 1}\ x_{i_q 2} \dots\ x_{i_q N}\right]^{\mathrm{T}}$, where $i_q$ is the range cell index the $q$-th training vector is associated with, $q = 1, 2, \dots, K_{\mathrm{T}}$, and $K_{\mathrm{T}}$ is the number of vectors. It is assumed that $\mathbf{x}_q$ are target-free, mutually, and share the same covariance matrix with the CUT vector $\mathbf{x}$ under hypothesis $H_0$ (target absent).

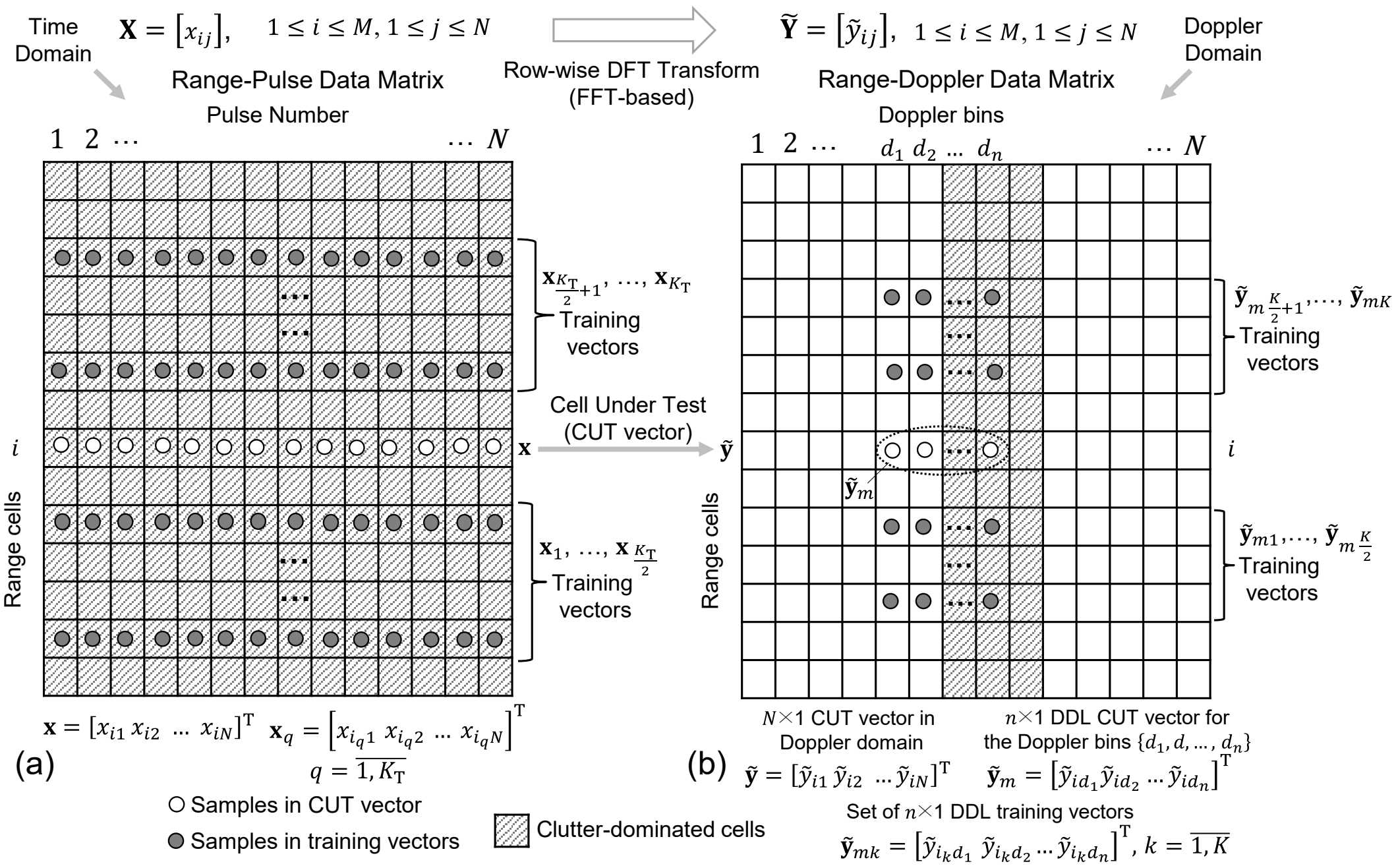

Fig. 1. Data arrangement comparison for adaptive detectors: (a) time domain and (b) Doppler domain

Fig. 1 (b) shows a similar data arrangement in the Doppler domain for adaptive target detection based on the DDL adaptivity principle (DDL principle for brevity). The range-Doppler data matrix $\tilde{\mathbf{Y}} = [\tilde{y}_{ij}]$, $M{\times}N$ results from the row-wise DFT applied to the data matrix $\mathbf{X}$ in Fig. 1 (a). In particular, the vector $\tilde{\mathbf{y}} = [\tilde{y}_{i1}\ \tilde{y}_{i2} \dots \tilde{y}_{iN}]^{\mathrm{T}}$ of size $N \times 1$ is a Doppler domain image of the corresponding time domain CUT vector $\mathbf{x} = [x_{i1} x_{i2} \dots x_{iN}]^{\mathrm{T}}$. The $n \times 1$ vector $\tilde{\mathbf{y}}_m = \left[\tilde{y}_{id_1} \tilde{y}_{id_2} \dots \tilde{y}_{id_n}\right]^{\mathrm{T}}$, $n \ll N$, is a DDL CUT vector to be tested by a DDL detector for the presence of an $m$-th presumable target with unknown Doppler frequency that may present in the $i$-th range cell. The vector $\tilde{\mathbf{y}}_m$ comprises the $n$ entries of the vector $\tilde{\mathbf{y}}$ corresponding to a set of Doppler bins $d_1, d_2, \dots, d_n$, where the essential portion of the $m$-th presumable target is concentrated. The procedure for determining such a set of Doppler bins will be introduced further. The shaded circles mark the samples in the training vectors $\tilde{\mathbf{y}}_{mk} = \left[\tilde{y}_{i_k d_1}\ \tilde{y}_{i_k d_2} \dots \tilde{y}_{i_k d_n}\right]^{\mathrm{T}}$, where $i_k$ is the range cell index the $k$-th training vector is associated with, $k = 1,2, \dots, K$, and $K$ ($K \ll K_{\mathrm{T}}$) is the number of vectors to be used for estimating an unknown DDL covariance matrix $\tilde{\mathbf{\Phi}}_m$ of size $n \times n$. The entries in the training vectors are associated with the Doppler bins $d_1, d_2, \dots, d_n$. The vectors $\tilde{\mathbf{y}}_{mk}$ are target-free, mutually independent, and share the same DDL covariance matrix $\tilde{\mathbf{\Phi}}_m$ with the DDL CUT vector $\tilde{\mathbf{y}}_m$ under hypothesis $H_0$.

It is straightforward to extend the theory of classical adaptive detection to the Doppler domain since Gaussian distributions are closed under affine transformations. This property means that for any complex-valued Gaussian vector $\mathbf{x}$, $\mathbf{x} \sim \mathcal{CN}(\boldsymbol{\mu}, \boldsymbol{\Sigma})$, a new vector $\tilde{\mathbf{y}} = \mathbf{B}\mathbf{x} + \mathbf{b}$ also follows a Gaussian distribution, $\tilde{\mathbf{y}} \sim \mathcal{CN}(\mathbf{B}\boldsymbol{\mu} + \mathbf{b}, \mathbf{B}\boldsymbol{\Sigma}\mathbf{B}^{\mathrm{H}})$, for all nonsingular $\mathbf{B} \in \mathbb{C}^{N \times N}$ and $\mathbf{b} \in \mathbb{C}^{N}$. The Discrete Fourier Transform (DFT) is a linear transformation given by $\tilde{\mathbf{y}} = \mathbf{F}\mathbf{x}$, with $\mathbf{F}$ being the DFT matrix. Hence, for a time domain vector $\mathbf{c} \sim \mathcal{CN}(\mathbf{0}, \boldsymbol{\Sigma})$, its Doppler domain image is a vector $\tilde{\mathbf{c}} \sim \mathcal{CN}(\mathbf{0}, \tilde{\boldsymbol{\Sigma}})$, with $\tilde{\boldsymbol{\Sigma}} = \mathbf{F}\boldsymbol{\Sigma}\mathbf{F}^{\mathrm{H}}$.

Thus, to build the theory of adaptive signal detection in the Doppler domain, one should follow the corresponding time domain theory and apply proper symbol substitutions for parameters, vectors, and matrices by their corresponding Doppler domain equivalents. This theory operates on DFT images of size $N\times1$ for vectors and $N\times N$ for matrices.

The theory of adaptive signal detection based on the DDL principle operates on vectors and matrices that represent Doppler domain data associated with a small set of Doppler bins $d_1, d_2, \dots, d_n, n \ll N$. Hence, the DDL-based theory of adaptive detection is a particular case of the adaptive detection theory in the Doppler domain.

We first define the Doppler domain representations (DFT images) corresponding to the time domain data. These are the $N\times1$ received data vector $\mathbf{z} = [z_1\ z_2 \dots z_N]^{\mathrm{T}}$, the $N\times1$ target steering vector $\mathbf{s} = [1\ e^{j2\pi F} \dots e^{j2\pi(N-1)F}]^{\mathrm{T}}$, $F$ is the known Doppler frequency, and the known disturbance covariance matrix $\mathbf{\Sigma}$ of size $N\times N$ (the term disturbance stands for noise plus clutter). This matrix is given by

$$\mathbf{\Sigma} = P_{\mathrm{c}}\mathbf{C}_0 + P_{\mathrm{n}}\mathbf{I} \tag{1}$$

where $P_{\mathrm{c}}$ and $P_{\mathrm{n}}$ represent the clutter and thermal noise power, $\mathbf{C}_0 = [c_{ij}]$ is the $N\times N$ normalized clutter covariance matrix ($c_{ii} = 1, i = 1,2, \dots, N$), and $\mathbf{I}$ is the $N\times N$ identity matrix.

In the Doppler domain, we use a zero-Doppler-centered data format. Therefore, the DFT images $\tilde{\mathbf{z}}$ and $\tilde{\mathbf{s}}$ of size $N\times1$ associated with the vector $\mathbf{z}$ and the vector $\mathbf{s}$ are given by

$$\tilde{\mathbf{z}} = [\tilde{z}_1\ \tilde{z}_2 \dots \tilde{z}_N]^{\mathrm{T}} = \mathrm{fftshift}(\mathrm{fft}(\mathbf{z})) \tag{2}$$

$$\tilde{\mathbf{s}} = [\tilde{s}_1\ \tilde{s}_2 \dots \tilde{s}_N]^{\mathrm{T}} = \mathrm{fftshift}(\mathrm{fft}(\mathbf{s})) \tag{3}$$

where fft and fftshift stand for Matlab functions, respectively implementing the FFT and the data shift needed to put the zero-frequency component in the middle of the DFT spectrum. Accordingly, the DFT image $\widetilde{\mathbf{\Sigma}}$, $N\times N$, corresponding to the disturbance covariance matrix $\mathbf{\Sigma}$ is given by

$$\widetilde{\mathbf{\Sigma}} = \mathrm{fftshift}(\mathrm{fft}(\mathbf{\Sigma})) \tag{4}$$

Having defined the DFT images of size $N\times1$ (vectors) and of size $N\times N$ (matrices), it is straightforward to determine the DDL representations for vectors and matrices associated with a given set of Doppler bins $d_1, d_2, \dots, d_n$. The $n\times1$ DDL steering vector $\tilde{\mathbf{t}} = [\tilde{t}_1\ \tilde{t}_2 \dots \tilde{t}_n]^{\mathrm{T}}$ associated with this set is generated by extracting from the $N\times1$ vector $\tilde{\mathbf{s}}$ the entries located at the $d_k$-th positions, i.e., $\tilde{t}_k = \tilde{s}_{d_k}$, $k = 1,2, \dots, n$. Similarly, the $n\times1$ DDL vector of received data $\tilde{\boldsymbol{\tau}} = [\tilde{\tau}_1\ \tilde{\tau}_2 \dots \tau_n]^{\mathrm{T}}$ is generated by extracting from the $N\times1$ vector $\tilde{\mathbf{z}}$ the entries located at the $d_k$-th positions, i.e., $\tilde{\tau}_k = \tilde{z}_{d_k}$, $k = 1,2, \dots, n$.

The $n\times n$ disturbance covariance matrix $\widetilde{\mathbf{\Phi}}$ associated with the set of Doppler bins $d_1, d_2, \dots, d_n$ is represented by the entries located at the intersection of the $d_k$-th rows and $d_l$-th columns, $k, l = 1,2,\ldots,n$ in the $N\times N$ matrix $\tilde{\mathbf{\Sigma}} = [\tilde{\sigma}_{mn}]$, $m, n = 1,2,\ldots,N$. Thus, the DDL covariance matrix $\tilde{\mathbf{\Phi}}$ is generated by extracting the corresponding entries in the matrix $\tilde{\mathbf{\Sigma}}$ as given below

$$\tilde{\mathbf{\Phi}} = [\tilde{\varphi}_{kl}],\ \tilde{\varphi}_{kl} = \tilde{\sigma}_{d_k d_l},\ k, l = 1,2,\ldots,n \tag{5}$$

To estimate the unknown covariance matrix $\mathbf{\Sigma}$ classical adaptive detectors employ a sample covariance matrix $\hat{\mathbf{\Sigma}}$ that is calculated using the set of training vectors $\mathbf{x}_q$, $q = 1, 2, \ldots, K_{\mathrm{T}}$ as

$$\hat{\mathbf{\Sigma}} = \frac{1}{K_{\mathrm{T}}} \sum_{q=1}^{K_{\mathrm{T}}} \mathbf{x}_q \mathbf{x}_q^{\mathrm{H}} \tag{6}$$

By analogy, an adaptive DDL detector of order $n$ employs the corresponding sample DDL covariance matrix $\hat{\tilde{\mathbf{\Phi}}}$ of size $n\times n$ as an estimate of the exact DDL covariance matrix $\tilde{\mathbf{\Phi}}$ associated with the set of Doppler bins $d_1, d_2, \ldots, d_n$. The sample DDL covariance matrix $\hat{\tilde{\mathbf{\Phi}}}$, $n\times n$, is calculated using the collection of the DDL training vectors $\tilde{\mathbf{y}}_{mk} = \left[\tilde{y}_{i_k d_1}\ \tilde{y}_{i_k d_2} \cdots \tilde{y}_{i_k d_n}\right]^{\mathrm{T}}$, $k = 1, 2, \ldots, K$ as

$$\hat{\tilde{\mathbf{\Phi}}} = \frac{1}{K} \sum_{k=1}^{K} \tilde{\mathbf{y}}_{mk} \tilde{\mathbf{y}}_{mk}^{\mathrm{H}} \tag{7}$$

## III. RODI-Based DDL-GLR Detector

As mentioned in Introduction, the RODI-based DDL-GLR detector [4] assumes that the entire Doppler domain may be divided into two types of regions. The first type refers to the flat or nearly flat spectrum regions, and the second one refers to the sharply changing spectrum regions. According to [4], if the target Doppler frequency is within the first type of spectrum regions and far away from the clutter spectrum peaks, the conventional processor employing a windowed FFT bank of filters followed by the traditional CA CFAR detectors is expected to provide nearly optimum detection performance (at least under Gaussian clutter). In contrast, when the target Doppler falls on sharply changing spectrum areas or regions near the clutter spectrum peaks, the traditional detector performance may fall far below the optimum. Therefore, [4] recommends using the adaptive DDL processing for improving detection performance only for such spectrum regions called the regions of detection improvement (RODI). The samples extracted from the Doppler bins in the $l$-th RODI $\{j_{l1}\ j_{l2} \ldots j_{ln_l}\}$ are used to form the corresponding $n_l \times 1$ CUT and training vectors to feed the DDL-GLR detector of the order $n_l$ associated with the $l$-th RODI. The order of the $l$-th RODI $n_l$ is the number of Doppler bins covered by this RODI, $n_l \ll N$.

Let $\tilde{\mathbf{y}}_l = [\tilde{y}_{ij_{l1}} \; \tilde{y}_{ij_{l2}} \dots \tilde{y}_{ij_{ln_l}}]^{\mathrm{T}}$ be the CUT vector, $n_l \times 1$, associated with the $l$-th RODI at the $i$-th range cell in the detection matrix $\tilde{\mathbf{Y}}$. Similarly, let $\tilde{\mathbf{y}}_{lk} = [\tilde{y}_{i_k j_{l1}} \; \tilde{y}_{i_k j_{l2}} \dots \tilde{y}_{i_k j_{ln_l}}]^{\mathrm{T}}$ be the $k$-th vector, with $i_k$ being the $k$-th range cell index the $k$-th training vector is associated with, $k = 1,2, \dots, K$ in a set of training data associated with the $\tilde{\mathbf{y}}_l$. Let $\tilde{\mathbf{s}}_l = [\tilde{s}_{j_{l1}} \; \tilde{s}_{j_{l2}} \dots \tilde{s}_{j_{ln_l}}]^{\mathrm{T}}$ be the $n_l \times 1$ target DDL steering vector (corresponding to some Doppler frequency), which is associated with the $l$-th RODI. The entries in $\tilde{\mathbf{s}}_l$ are extracted from the positions $j_{lm}, m = 1,2, \dots, n_l$ in the $N \times 1$ target steering vector in the Doppler domain

$$\tilde{\mathbf{s}} = [\tilde{s}_1 \; \tilde{s}_2 \dots \tilde{s}_{j_{l1}} \; \tilde{s}_{j_{l2}} \dots \tilde{s}_{j_{ln_l}} \dots \tilde{s}_{N-1} \; \tilde{s}_N]^{\mathrm{T}} \tag{8}$$

The vector $\tilde{\mathbf{s}}$ in (8) is the result of the FFT transform (unweighted or weighted) applied to the $N \times 1$ target steering vector in the time domain $\mathbf{s} = [1 \; e^{j2\pi F} \; e^{j2\pi 2F} \dots e^{j2\pi(N-1)F}]^{\mathrm{T}}$, where $F$ is the normalized target Doppler frequency, $|F| < 0.5$.

For the $l$-th RODI $\{j_{l1} \; j_{l2} \dots j_{ln_l}\}$, the DDL-GLR detector of the order $n_l$ computes $n_l$ test statistics $\eta_{lm}$ covering the Doppler bins $j_{lm}, m = 1,2, \dots, n_l$, as given by

$$\eta_{lm} = \frac{1}{\tilde{\mathbf{s}}_{lm}^{\mathrm{H}} \hat{\boldsymbol{\Sigma}}_l^{-1} \mathbf{s}_{lm}} \frac{|\tilde{\mathbf{s}}_{lm}^{\mathrm{H}} \hat{\boldsymbol{\Sigma}}_l^{-1} \tilde{\mathbf{y}}_l|^2}{1 + (\tilde{\mathbf{y}}_l^{\mathrm{H}} \hat{\boldsymbol{\Sigma}}_l^{-1} \tilde{\mathbf{y}}_l)/K} \tag{9}$$

where $\hat{\boldsymbol{\Sigma}}_l = (1/K) \sum_{k=1}^{K} \tilde{\mathbf{y}}_{lk} \tilde{\mathbf{y}}_{lk}^{\mathrm{H}}$, $n_l \times n_l$, is the interference covariance matrix estimate associated with the $l$-th RODI.

In (9) above, the $m$-th vector $\tilde{\mathbf{s}}_{lm} = [\tilde{s}_{j_{l1}} \; \tilde{s}_{j_{l2}} \dots \tilde{s}_{j_{ln_l}}]^{\mathrm{T}} = [\tilde{s}_{1m} \; \tilde{s}_{2m} \dots \tilde{s}_{n_l m}]^{\mathrm{T}}, m = 1,2, \dots, n_l$, is the DDL steering vector associated with the $m$-th Doppler bin in the $l$-th RODI. The entries in $\tilde{\mathbf{s}}_{lm}$ are extracted from the vector $\tilde{\mathbf{s}}$ as specified by (8). For each $m$, the vector $\tilde{\mathbf{s}}$ in (8) is the result of the FFT applied to the time domain vector $\mathbf{s}$ computed for the on-grid Doppler frequency $F = F_{lm} = \frac{j_{lm}}{N}$ associated with the $m$-th Doppler bin of the $l$-th RODI. Hence, no calculations are needed to determine DDL steering vectors associated with Doppler bins: all entries in the $m$-th vector $\tilde{\mathbf{s}}_{lm}, m = 1,2, \dots, n_l$ are zeros except the $lm$-th one, which is unity.

Each test statistic $\eta_{lm}, m = 1,2, \dots, n_l$ in (9) is then compared with the threshold $\eta_l$ to complete the hypothesis test for the $l$-th RODI. The tests for Doppler bins not in any RODI can be performed by some conventional CFAR method. The false alarm probability of the DDL-GLR at any bin of the $l$-th RODI relates to the detection threshold $\eta_l$ as given by

$$P_{\mathrm{FA}l} = \left(1 - \frac{\eta_l}{n_l}\right)^{K - n_l + 1} \tag{10}$$

It should be noted that the GLR statistic $\eta_{lm}$ in (9) assumes that the target signal does not contain unknown parameters. In [1], Kelly pointed out that if the signal steering vector includes one or more unknown parameters (such as target Doppler), the test statistic has to be maximized over these parameters. The maximization over target Doppler generally cannot be carried out explicitly. A remedy to this problem is a standard technique [1]: approximating the test statistic by computing it on a discrete set of Doppler frequencies. This technique is equivalent to forming a filter bank and declaring target presence if any filter output exceeds the threshold.

Reference [4] analyzes the detection performance of the RODI-based DDL-GLR detector only for the non-fluctuating target model (the target signal amplitude is a constant) with a known target Doppler. This detector may provide reliable target detection under these assumptions even when the RODI order $n_l$ is essentially smaller than $N$. Presented numerical results demonstrate that even for $n_l = 4 - 6$, it may achieve near-optimum detection performance delivered by the optimum detector of the order $N = 64$.

An example discussed in [4, p. 535, Fig. 2] shows the RODI-based DDL-GLR detector's performance on the DFT grid for the known target Doppler. The strong clutter is assumed to have a Gaussian-shaped PSD. The detector employs two adjacent RODIs of the 4-th order placed next to the PSD peak; these RODIs cover the Doppler bins from 11 to 18 (non-zero-Doppler-centered axis). The total number of pulses in the CPI is $N = 64$, with $K = 24$ available training samples. The clutter PSD is centered at the normalized Doppler frequency $F_{\text{cp}} = 0.15$ (peak location) with the spread parameter $\sigma_{\text{c}} = 0.0025$. The clutter-to-noise ratio (CNR) is set to 60 dB. The signal-to-noise ratio (SNR) is 3 dB, which results in the signal-to-disturbance ratio SDR = –57 dB at the detector's input. This example shows near-optimum detection for the DDL-GLR detector at each RODI-associated Doppler bin, except for a slight departure from the optimum at bin 15.

An essential performance departure from the optimum at both on-grid and off-grid Doppler frequencies is demonstrated by the frequency response ($P_{\text{D}}$ versus Doppler frequency) for the 4th order RODI-based DDL-GLR detector [see 4, p. 536, Fig. 3]. Analyzing the frequency response shows that there is a need to improve the detection performance around some off-grid frequency point ("bad" frequency point). To remedy this problem, [4] suggests using an additional target steering vector corresponding to the middle of the nearest on-grid points. However, the radar processor needs accurate information on these "bad" frequency points to apply this remedy. In real radar systems, such information is unavailable and can only be obtained by measurements from the received data. These measurements complicate the RODI-based DDL-GLR implementation, especially in multiple clutter situations. In turn, measurement errors will inevitably result in performance degradation.

This section complements the results in [4] with the detection performance analysis for the Swerling I target model, assuming a known target Doppler. We will show that the RODI-based DDL-GLR detector may suffer essential detection performance departures from the optimum. These departures may occur even for the known target Doppler frequency, no matter whether it takes on or off DFT grid values. The said detection performance degradations prevent implementing the RODI-based DDL-GLR detector under unknown Doppler frequency by using DDL-GLR test statistics associated with the DFT grid.

For the known target Doppler frequency, the time domain GLR and AMF (TD-GLR and TD-AMF) detectors' equations describing the detection performance for the Swerling I target model are given in [2]. Following [2], Appendices B and C derive similar DDL-GLR and DDL-AMF detectors' equations. The corresponding equation for the optimum detector of order $N$ is given in Appendix A.

Fig. 2 shows the on-grid detection performance of the RODI-based DDL-GLR detector as $P_{\mathrm{D}}$-vs-$F$ plots (probability of detection $P_{\mathrm{D}}$ versus normalized Doppler frequency $F$). In this figure, the number of pulses and training samples equals those in [4]: $N = 64$ and $K = 24$, respectively. Other settings are the Gaussian-shaped clutter PSD centered at $F_{\mathrm{cp}} = 0.15$ or $F_{\mathrm{cp}} = 0.156$ with $\sigma_{\mathrm{c}} = 0.0025$, CNR = 60 dB, and SNR = 10 dB, which gives the input SDR = –50 dB. The two adjacent RODIs of the 4-th order next to the clutter PSD peak are $\mathrm{RODI}_1$ = [43, 44, 45, 46] and $\mathrm{RODI}_2$ = [47, 48, 49, 50] (we use a zero-Doppler-centered axis).

As seen in Fig. 2 (a), the $P_{\mathrm{D}}$ value ($P_{\mathrm{D}} = 0.7536$) at the Doppler bin 47 is noticeably below the optimum, while it keeps near the optimum for the neighboring Doppler bins. Fig. 2 (b) illustrates the performance degradation for $F_{\mathrm{cp}} = 0.156$; this $F_{\mathrm{cp}}$ value is just a 4% shift relative to $F_{\mathrm{cp}} = 0.15$ in Fig. 2 (a). Comparing Fig. 2 (a) and Fig. 2 (b) shows that the performance of the RODI-based DDL-GLR detector is quite sensitive to the location of clutter PSD peak. This sensitivity is especially pronounced in Doppler regions with a sharp change in the clutter spectrum. Moreover, it will also result in inevitable performance degradation due to errors in measuring clutter spectrum parameters.

Fig. 3 shows the probability of detection $P_{\mathrm{D}}$ at the three off-grid Doppler frequencies for the DDL-GLR detector associated with $\mathrm{RODI}_1$ and $\mathrm{RODI}_2$ previously used in Fig. 2. These off-grid frequencies are $F_1 = \frac{13}{N} + \frac{0.25}{N} = 0.2070$, $F_2 = \frac{13}{N} + \frac{0.5056}{N} = 0.2110$, and $F_3 = \frac{13}{N} + \frac{0.75}{N} = 0.2148$. As can be seen, there are essential departures from the optimum at the off-grid frequencies $F_2$ and $F_3$.

To explain this effect, we consider the signal power distribution across the Doppler bins in the corresponding RODIs for each of these specific Doppler frequencies. Figs. 4 (a) through (d) respectively plot the normalized (unit norm) DFT power spectra for the target signals whose Doppler frequencies are $F_{46}$, $F_1$, $F_2$, and $F_3$ - analyzing these spectra yields that the $P_{\mathrm{D}}$ value approaches closer to the optimum when the portion of the total signal power in the corresponding RODI becomes more significant. In Fig. 4 (a), for the signal at frequency $F_{46} = 13/N$, the total signal power is in the RODI$_1$= [43, 44, 45, 46] and, correspondingly, the associated $P_\mathrm{D} = 0.8796$ is close to the optimum.

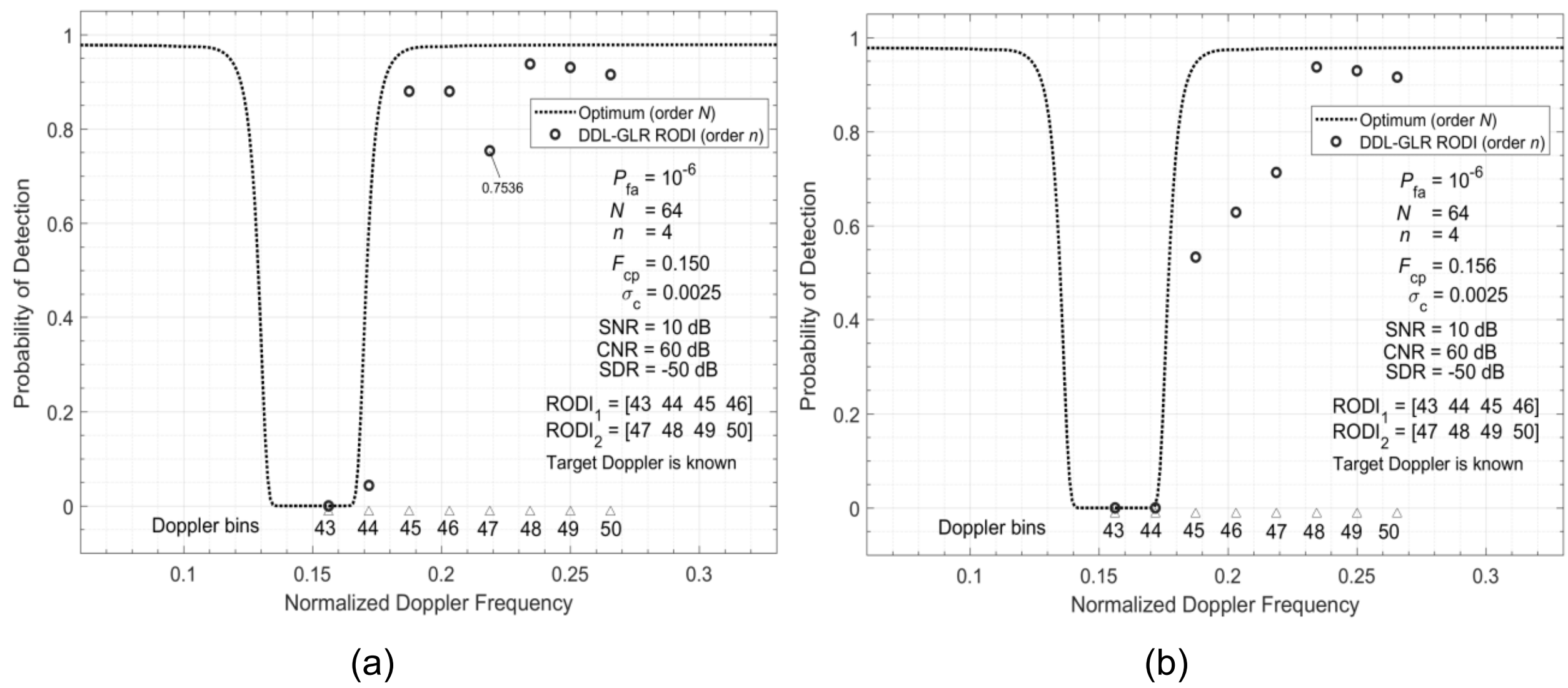

Fig. 2. On-grid detection performance of RODI-based DDL-GLR detector for Swerling I target

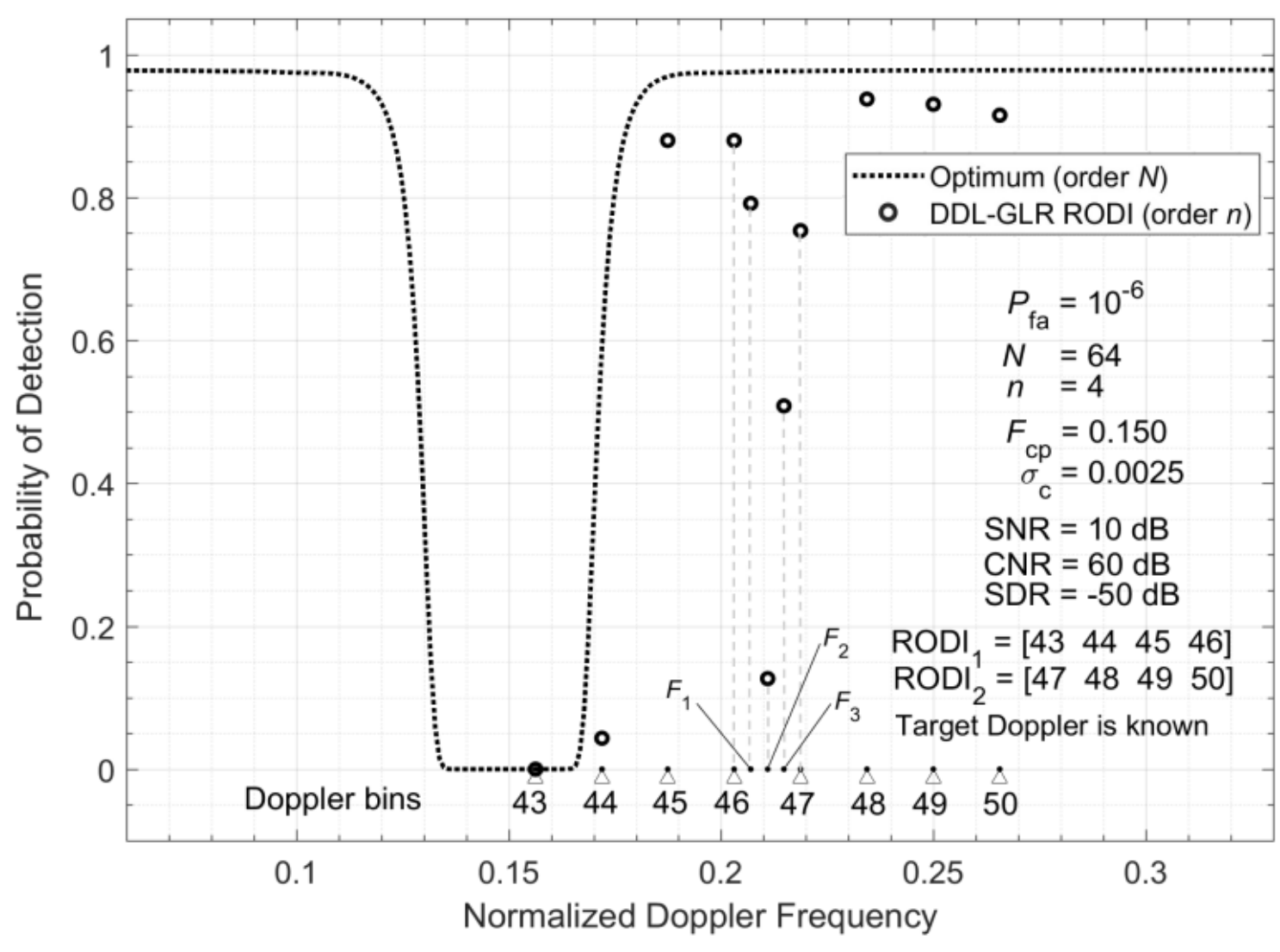

Fig. 3. On- and off-grid detection performance of RODI-based DDL-GLR detector for Swerling I target

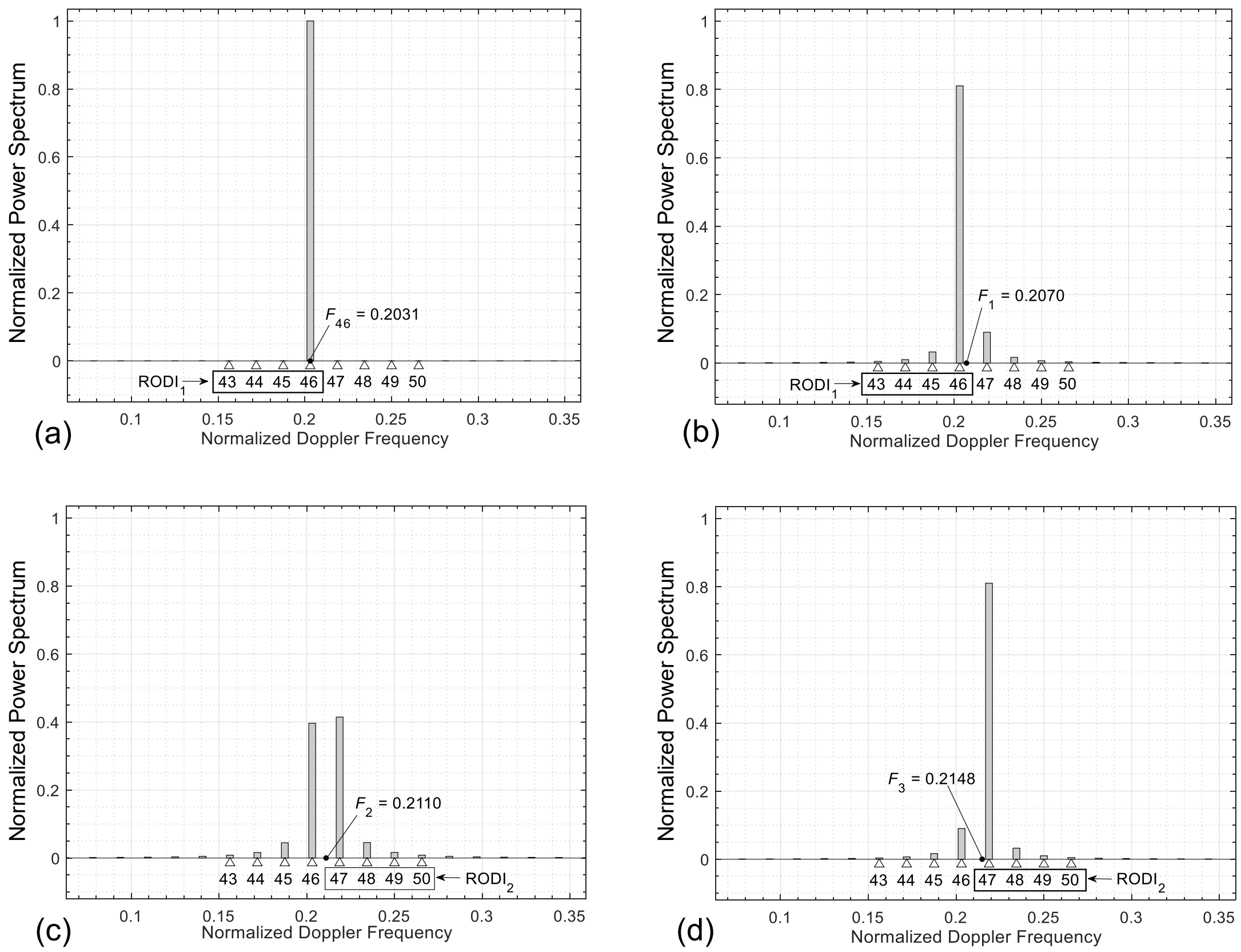

Fig. 4. Target signal power distribution across Doppler bins in RODIs

In Fig. 4 (b), for the signal at frequency $F_1 = 0.2070$, the portion of its total power in the RODI$_1$=[43, 44, 45, 46] exceeds 0.8, and the corresponding $P_{\mathrm{D}} = 0.7916$ is not too far from the optimum. In contrast, in Fig. 4 (c), for the signal at $F_2 = 0.2110$, the portion of its total power in the RODI$_2$=[47, 48, 49, 50] is about 0.5, and the corresponding $P_{\mathrm{D}} = 0.1268$ is far from the optimum. For the signal at $F_3 = 0.2110$, the portion of its total power in the RODI$_2$ is equal to that for the signal at $F_1$ in the RODI$_1$. However, the $P_{\mathrm{D}}$ value at $F_3$ is less than that at $F_1$. This is because the clutter samples in RODI$_2$ are less correlated than in RODI$_1$; therefore, the clutter associated with the RODI$_2$ is less suppressed than that in RODI$_1$.

Inspecting Fig. 4 (c) and (d) leads to the conclusion that choosing new RODI$_2$ = [45, 46, 47, 48] would bring the $P_{\mathrm{D}}$ values at frequencies $F_2$ and $F_3$ near the optimum. However, this choice is based on the knowledge of clutter PSD parameters. Unfortunately, they are unknown in the actual radar operational environment.

As pointed out in [4], determining the number and locations of RODIs required to achieve near-optimum detection performance at arbitrary target Doppler frequency is a challenging problem.

However, [4] suggests no practical solutions to this problem; moreover, we have not found answers in the open literature.

Analyzing the effect of the signal power distribution across RODI on DDL-GLR performance yields that using the RODI concept in DDL processing does not maximize the signal power portion within RODI as required to achieve near-optimum detection at arbitrary target Doppler frequency. Thus, the shortcomings of the RODI-based DDL-GLR detector [4] make its practical use questionable.

## IV. Region of Possible Target Detection

The analysis of the signal power distribution within a limited Doppler region on the DDL-GLR detection performance reveals that the DDL detector has to capture as much signal power as possible to have its detection performance closer to the optimum. This observation also suggests that determining the limited Doppler region for a DDL detector must be based on the target signal spectrum rather than the clutter spectrum to capture the maximum possible portion of the target signal power. This idea leads to the *Region of Possible Target Detection* (RPTD) concept. As shown below, the RPTD-based DDL detectors can provide near-optimum detection performance at arbitrary target Doppler, no matter whether it is known or not.

For a DDL detector of order $n$, the region of possible target detection (RPTD) is such a set of $n$ Doppler bins $\mathcal{D} = \{d_1, d_2, \dots, d_n\}$ that with high probability contains the maximal portion of the power of a target signal that may presumably be present in the cell under test associated with a given range cell in the range-Doppler data matrix (or the detection matrix) $\widetilde{\mathbf{Y}}$.

As is well known, the DFT concentrates an essential portion of the signal power within some narrow Doppler region around the peak value in the signal power distribution. Fig. 5 (b) illustrates the power distribution for clutter and target signals in the matrix $\tilde{\mathbf{Z}} = [\tilde{z}_{ij}] = [|\tilde{y}_{ij}|^2]$, $M \times N$, calculated from the corresponding entries in the complex-valued detection matrix $\widetilde{\mathbf{Y}} = [\tilde{y}_{ij}]$, $M \times N$. The detection matrix $\widetilde{\mathbf{Y}}$ is the result of the row-wise FFT transform applied to the original CPI data matrix shown in Fig. 5 (a). For simplicity, Fig. 5 (a) shows only one target-associated sample (maximum magnitude) on the range axis. In Fig. 5 (b), for each range cell that contains target signal data, we call the region of signal power concentration (SPC) a set of Doppler bins $SPC = \{d_1, d_2, \dots, d_n\}$, $n \ll N$, where some essential portion of the signal power is confined.

Similarly, we denote $CPC = \{d_1, d_2, \dots, d_m\}$, $m \ll N$, the region of clutter power concentration. Fig. 5(b) shows two isolated regions $SPC_{i_1} = \{d_7\ d_8\ d_9\}$ and $CPC_{i_1} = \{d_3\ d_4\ d_5\}$, at the $i_1$-th range bin, and two overlapping regions $SPC_{i_2} = \{d_2\ d_3\ \ d_4\}$ and $CPC_{i_2} = \{d_3\ d_4\ d_5\}$, at the $i_2$-th range bin. Since SPC and CPC regions may overlap, the data associated with CPC regions may also contain the target signal power.

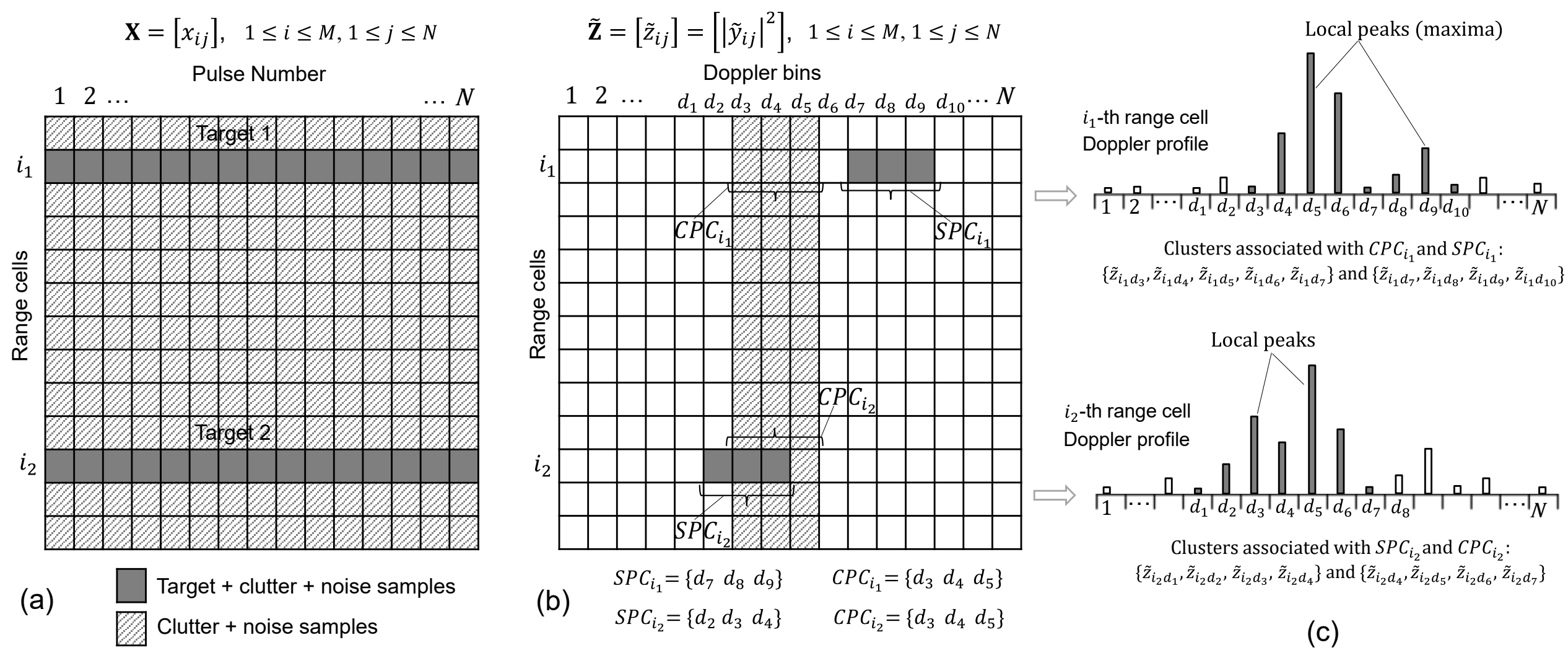

Fig. 5. SPC and CPC regions and associated with them clusters in Doppler profiles

Fig. 5 (c) illustrates the Doppler profiles associated with the range cells that contain SPC and CPC regions depicted in Fig. 5 (b). These Doppler profiles represent magnitude-squared data samples from the corresponding rows in the matrix $\tilde{\mathbf{Z}}$. The data samples in each Doppler profile can be grouped in clusters. We define a cluster as a group of samples consisting of the maximum sample (local peak) and its neighboring samples up to the nearest minimum sample on the right and left of the local peak. Among all possible clusters in a Doppler profile, clusters associated with SPC and CPC regions are power-dominant since the signal-to-noise ratio (SNR) and clutter-to-noise ratio (CNR) increase after pulse compression and coherent integration (FFT). Samples in such SPC/CPC-associated clusters are shadowed in Fig. 5 (c). However, which clusters are due to targets is unknown a priori. Therefore, the data in the detection matrix $\tilde{\mathbf{Y}}$ associated with each cluster must be tested for the target presence.

The local peak value in a target-associated cluster with a high probability is the closest to the maximum value of the target signal power spectrum. Hence, to determine the RPTD of order $n$, we need to locate the peak value in a cluster and select a group of $n-1$ bins closest to the peak-associated Doppler bin to maximize the portion of the signal power captured by this group.

To introduce the RPTD identification procedure, we assume that the target Doppler $F$ is known ($|F| < 0.5$). This procedure comprises the following steps. First step computes the target steering vector $\mathbf{s} = [1\ e^{j2\pi F} \dots e^{j2\pi(N-1)F}]^{\mathrm{T}}$, its DFT image $\tilde{\mathbf{s}} = [\tilde{s}_1\ \tilde{s}_2 \dots \tilde{s}_N]^{\mathrm{T}} = \mathrm{fftshift}(\mathrm{fft}(\mathbf{s}))$, and the corresponding Doppler profile $\tilde{\mathbf{p}}$

$$\tilde{\mathbf{p}} = [\tilde{p}_1\ \tilde{p}_2\ \dots\ \tilde{p}_N],\ \tilde{p}_i = |\tilde{s}_i|^2, i = 1,2,\dots,N \tag{11}$$

The vector $\tilde{\mathbf{p}}$ represents the steering vector's power distribution in the Doppler axis. The second step finds the Doppler index $d_m$ associated with the maximum value $\tilde{p}_{d_m}$ in $\tilde{\mathbf{p}}$

$$\begin{gathered} d_m = \arg\max_i(\tilde{p}_i\,, i = 1,2,\dots,N) \\ \tilde{p}_{d_m} = \max([\tilde{p}_1\ \tilde{p}_2\ \dots\ \tilde{p}_N]) \end{gathered} \tag{12}$$

In the second step, the procedure may be essentially simplified. One can find $d_m$ using the simple equation $d_{\mathrm{o}} = \mathrm{round}(FN) + N/2 + 1$, where the function $\mathrm{round}(x)$ returns the nearest integer to $x$. The value of $d_{\mathrm{o}}$ needs correction considering the DFT periodicity $-$ if $d_{\mathrm{o}} < 1$, then $d_m = d_{\mathrm{o}} + N$; if $d_{\mathrm{o}} > N$, then $d_m = d_{\mathrm{o}} - N$; otherwise, $d_m = d_{\mathrm{o}}$.

Finally, the third step computes the RPTD associated set $\mathcal{D} = \{d_1\ d_2 \dots d_n\}$ using an algorithm shown in Fig. 6. This algorithm uses two closest neighboring samples $\tilde{p}_{d^-}$ and $\tilde{p}_{d^+}$ computed considering the DFT spectrum periodicity from the following equations

$$\begin{gathered} d^- = \begin{cases} a, & if\ a > 0 \\ N, & \text{otherwise} \end{cases} \quad \text{where } a = d_m - 1 \\ d^+ = \begin{cases} b, & if\ b \le N \\ 1, & \text{otherwise} \end{cases} \quad \text{where } b = d_m + 1 \end{gathered} \tag{13}$$

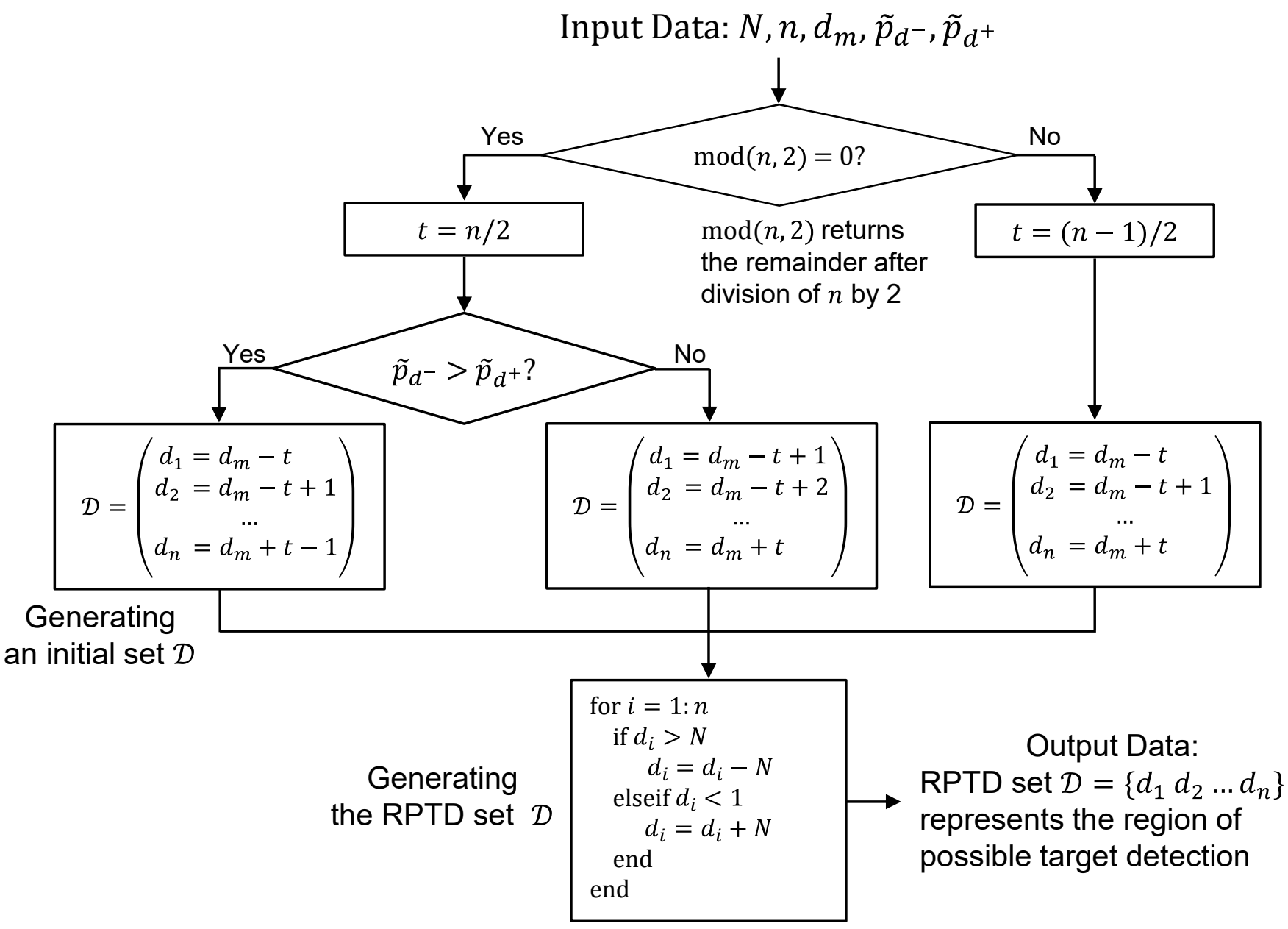

Fig. 6. RPTD computing algorithm

The RPTD identification procedure described above can also be used when the target Doppler frequency is unknown. In this case, one should substitute an unknown value of the target Doppler $F$ with its estimate $\hat{F}$. This estimate can be obtained using fine Doppler estimators [22].

## V. RPTD-Based DDL detectors

In the case of an unknown target Doppler frequency, the standard technique is to test the outputs of a filter bank associated with a fixed DFT grid. For the adaptive DDL processing, the present paper proposes a non-uniform grid whose elements are Doppler frequency estimates associated with local peaks (maxima) in a given Doppler profile. We refer to the Doppler grid defined in such a way as the maximum likelihood Doppler (MLD) grid.

Before calculating the MLD grid, we generate a set of representative range cells. To this end, we use the data in Doppler lines which are the columns $[\tilde{z}_{1j}\ \tilde{z}_{2j}\ \dots\ \tilde{z}_{Mj}]^{\mathrm{T}}$, $j = 1,2,\dots,N$, of the matrix $\tilde{\mathbf{Z}} = [\tilde{z}_{ij}] = [|\tilde{y}_{ij}|^2]$, $M{\times}N$. First, we find the positions of all local range peaks (maxima) in each Doppler line of the matrix $\tilde{\mathbf{Z}}$. For example, this location procedure can be done using the Matlab function "findpeaks." Then we compose a comprehensive set of representative range cell indices as the set $\mathcal{R} = \{i_k, k = 1,2,\dots,M_{\mathrm{R}}\}$, where the unique $M_{\mathrm{R}}$ indices $i_k \in \{1,2,\dots,M\}$ , $M_{\mathrm{R}} < M$, are all those range cell indices for which there exists a local range peak at least in one out of the $N$ Doppler lines.

Having obtained the set $\mathcal{R}$, we compute power Doppler profiles associated with the representative range cells. This is done by applying an $N_{\mathrm{fft}}$-point FFT to each vector $\mathbf{x} = [x_{i_k1}\ x_{i_k2}\ \dots\ x_{i_kN}]$ with the subsequent elementwise magnitude-squaring. The vectors $\mathbf{x}$ are extracted from the $i_k$-th row of the CPI data matrix $\mathbf{X}$, where $i_k \in \mathcal{R}$. In computing these profiles, we use the $N_{\mathrm{fft}}$-point FFT with $N_{\mathrm{fft}} = qN$, $q \geq 2$ because it provides favorable initial conditions for accurate Doppler measurements with a simple non-iterative fine Doppler estimator operating on magnitude data [22]. This computationally efficient estimator uses a peak and two neighboring samples for each local maximum in a given Doppler profile. In addition, such $qN$-point FFT avoids gross measurement errors at Doppler frequencies near ±0.5.

Let $[\tilde{v}_{i_k1}\ \tilde{v}_{i_k2}\ \dots\ \tilde{v}_{i_kN_{\mathrm{fft}}}]$ be a power Doppler profile associated with the $i_k$-th representative range cell, $k = 1,2,\dots,M_{\mathrm{R}}$. To obtain the MLD grid, we first find the positions of all local peaks in the vector $[\tilde{v}_1\ \tilde{v}_2\ \dots\ \tilde{v}_{N_{\mathrm{fft}}}]$ representing this Doppler profile (for simplicity, the range cell index $i_k$ for the samples $\tilde{v}_{i_kj}$ will be omitted). Since we have to consider the periodicity of the Doppler profile, a local peak identification procedure uses a $1{\times}(N_{\mathrm{fft}} + 2)$ vector $[\tilde{v}_{N_{\mathrm{fft}}}\ \tilde{v}_1\ \tilde{v}_2\ \dots\ \tilde{v}_{N_{\mathrm{fft}}}\ \tilde{v}_1]$ that is a given Doppler profile vector appended by the element $\tilde{v}_{N_{\mathrm{fft}}}$ on the left and the element $\tilde{v}_1$ on the right.

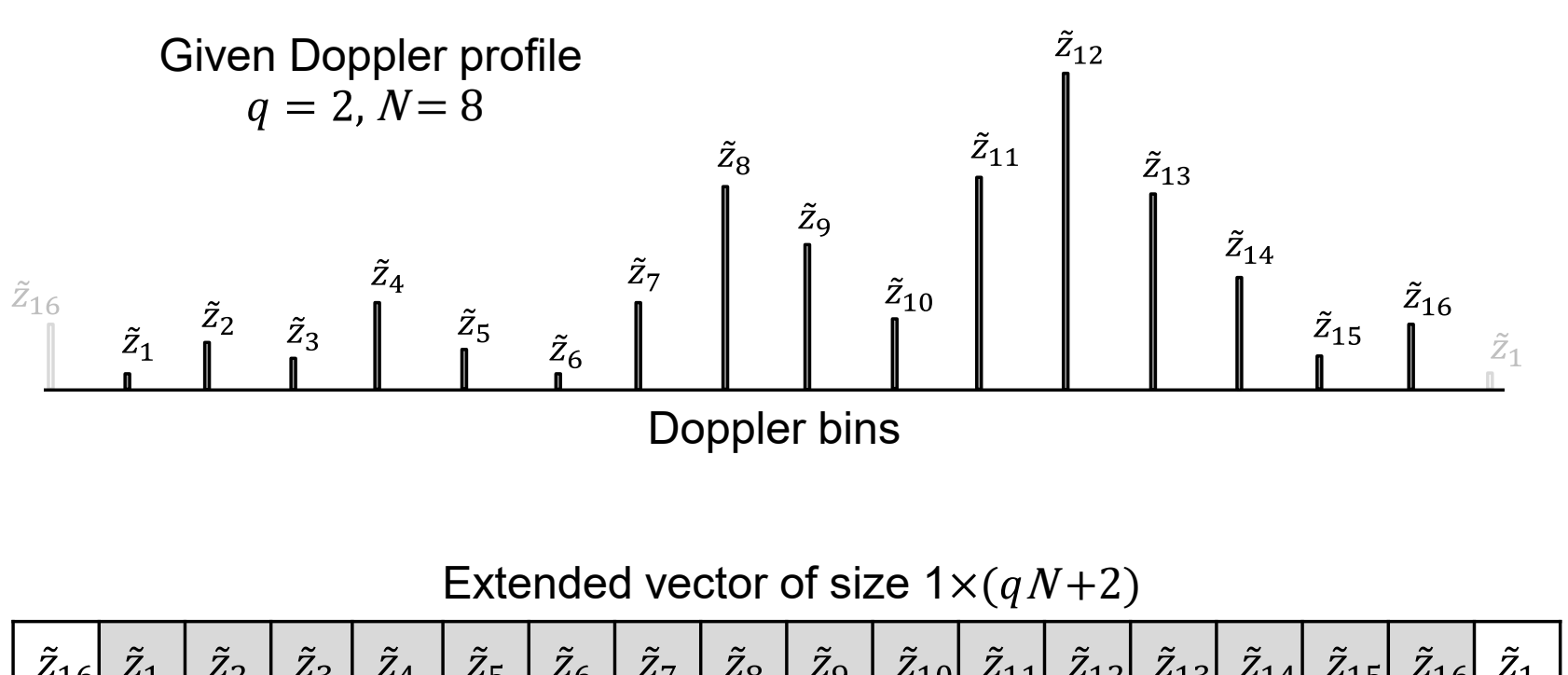

Extended vector of size $1 \times (qN+2)$

| $\tilde{z}_{16}$ | $\tilde{z}_1$ | $\tilde{z}_2$ | $\tilde{z}_3$ | $\tilde{z}_4$ | $\tilde{z}_5$ | $\tilde{z}_6$ | $\tilde{z}_7$ | $\tilde{z}_8$ | $\tilde{z}_9$ | $\tilde{z}_{10}$ | $\tilde{z}_{11}$ | $\tilde{z}_{12}$ | $\tilde{z}_{13}$ | $\tilde{z}_{14}$ | $\tilde{z}_{15}$ | $\tilde{z}_{16}$ | $\tilde{z}_1$ |
|---|---|---|---|---|---|---|---|---|---|---|---|---|---|---|---|---|---|
| $v_1$ | $v_2$ | $v_3$ | $v_4$ | $v_5$ | $v_6$ | $v_7$ | $v_8$ | $v_9$ | $v_{10}$ | $v_{11}$ | $v_{12}$ | $v_{13}$ | $v_{14}$ | $v_{15}$ | $v_{16}$ | $v_{17}$ | $v_{18}$ |

Output of "findpeaks" algorithm for given Doppler profile

The number of local maxima $k = 5$

The set of their positions $\{2, 4, 8, 12, 16\}$

Fig. 7. Local peaks identification procedure ("findpeaks" is a Matlab function)

Fig. 7 explains the simple procedure for identifying local peaks and shows its output for the profile of length 16 ($q = 2, N = 8$). Having obtained the positions $\{j_{km}, m = 1,2, \dots, P_k\}$ of $P_k$ local peaks in a power Doppler profile associated with the $i_k$-th range cell, $i_k \in \mathcal{R}$, we calculate the corresponding MLD grid by applying a fine Doppler estimator to each data set associated with the $m$-th local peak, $m = 1,2, \dots, P_k$. A simple fine Doppler estimator [22] uses a closed-form equation based on quadratic interpolation. This estimator measures the Doppler frequency estimate $\hat{f}_m$ as the peak location of the interpolating parabola fitted through the $m$-th local peak $\tilde{z}_{j_m}$ and two neighboring samples $\tilde{z}_{j^-}$ and $\tilde{z}_{j^+}$. The normalized Doppler frequency estimate is calculated as $\hat{F}_m = \hat{f}_m/(N_{\text{fft}})$.

It should be noted that some of the local peaks in a Doppler profile corresponding to the $i_k$-th representative range cell, $i_k \in \mathcal{R}$ represent range-Doppler peaks $\mathcal{P}_m(i_k, j_{km})$ associated with two-dimensional (range-Doppler) point target responses. These responses are due to radar returns from individual specular points (scattering centers) of various objects within a radar scene represented in the detection matrix $\widetilde{\mathbf{Y}}$. Each point target response consists of a group of correlated samples. Some of these groups are due to targets of interest that may present among the objects in the radar scene. Therefore, the data in such groups are of primary interest for detection hypothesis testing.

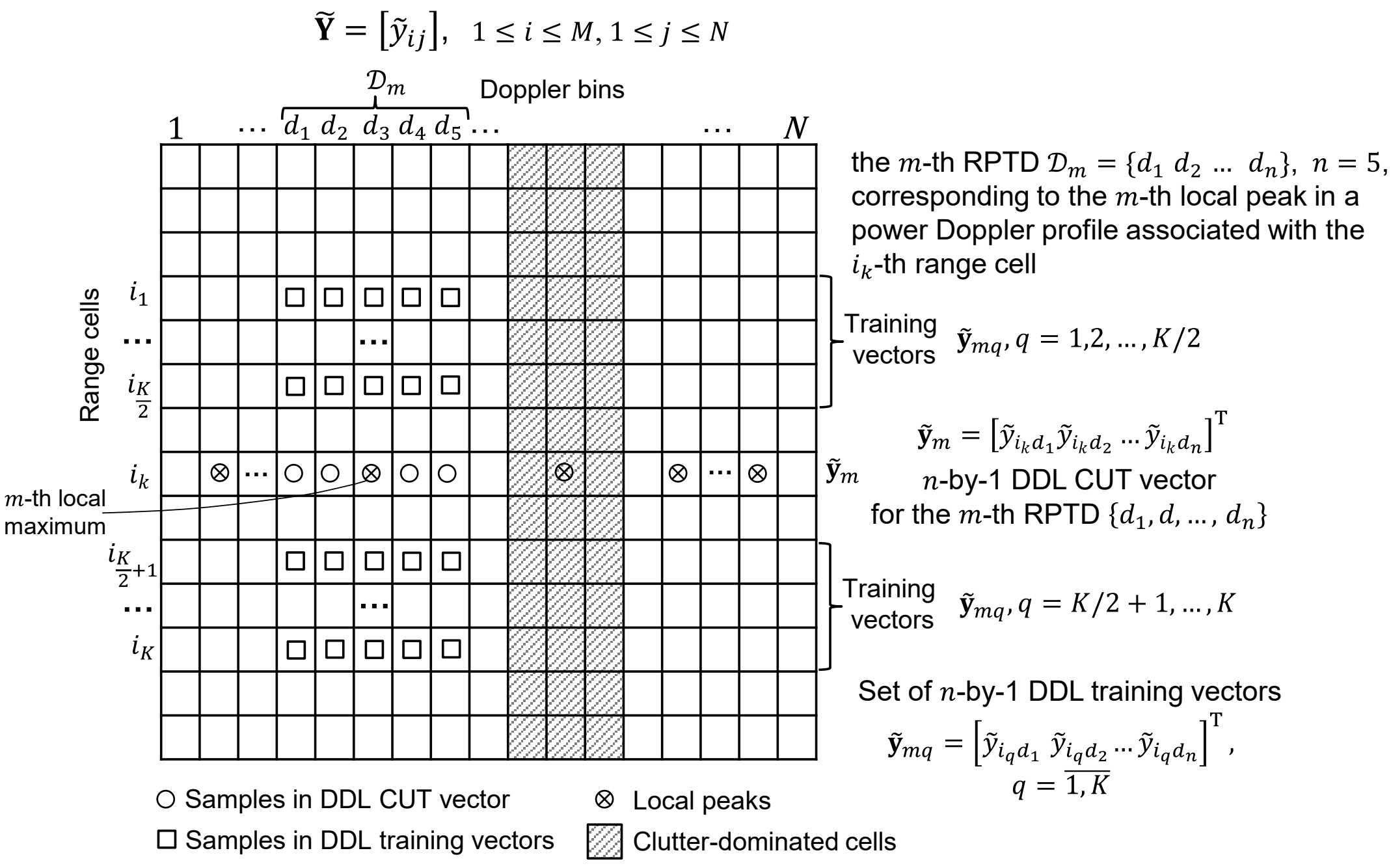

Fig. 8. CUT and training data for RPTD-based DDL detector

We use the Doppler frequency estimates $\hat{F}_m$, $m = 1,2,\dots,P_k$, from the MLD grid associated with the $i_k$-th range cell, $i_k \in \mathcal{R}$, with $P_k$ being the number of local peaks, in determining the $m$-th RPTD $\mathcal{D}_m = \{d_1, d_2, \dots, d_n\}$ associated with the $m$-th local Doppler peak. The elements in the set $\mathcal{D}_m$ are the indices of columns in the detection matrix $\tilde{\mathbf{Y}}$, from which the Doppler data are extracted for testing the target presence in the $m$-th RPTD.

Fig. 8 shows the location of the DDL data (CUT and training vectors) in the range-Doppler detection matrix $\tilde{\mathbf{Y}}$. The $n \times 1$ vector $\tilde{\mathbf{y}}_m = [\tilde{y}_{i_kd_1}\ \ \tilde{y}_{i_kd_2} \cdots \tilde{y}_{i_kd_n}]^{\mathrm{T}}$ is the DDL CUT vector formed by the entries at the columns associated with the $m$-th RPTD $\mathcal{D}_m = \{d_1, d_2, \dots, d_n\}$, ($n \ll N$) in the $i_k$-th row of the matrix $\tilde{\mathbf{Y}}$. Similarly, the $n \times 1$ vector $\tilde{\mathbf{y}}_{mq} = [\tilde{y}_{i_qd_1}\ \tilde{y}_{i_qd_2} \cdots \tilde{y}_{i_qd_n}]^{\mathrm{T}}$ is the $q$-th DDL training vector ($q = 1,2,\dots,K$) associated with the same $m$-th RPTD $\mathcal{D}_m$. The data for the $q$-th training vector are extracted from the corresponding $i_q$-th row of $\tilde{\mathbf{Y}}$.

Let $\hat{\tilde{\mathbf{\Phi}}}_m$ be the estimate of the exact DDL covariance matrix $\tilde{\mathbf{\Phi}}_m$ associated with the $m$-th RPTD $\mathcal{D}_m$. We compute this estimate as the sample DDL covariance matrix

$$\hat{\tilde{\mathbf{\Phi}}}_m = \frac{1}{K}\sum_{q=1}^{K}\tilde{\mathbf{y}}_{mq}\tilde{\mathbf{y}}_{mq}^{\mathrm{H}} \tag{14}$$

It is assumed that $K > n$; hence, the matrix $\hat{\tilde{\mathbf{\Phi}}}_m$ is not singular, so its inverse $\hat{\tilde{\mathbf{\Phi}}}_m^{-1}$ exists.

Let $\hat{\tilde{\mathbf{t}}}_m = [\hat{\tilde{t}}_1 \; \hat{\tilde{t}}_2 \; \dots \; \hat{\tilde{t}}_n]^{\mathrm{T}}$ be the DDL steering vector estimate formed respectively by the entries $\hat{\tilde{s}}_{d_1}, \hat{\tilde{s}}_{d_2}, \dots, \hat{\tilde{s}}_{d_n}$ that are extracted from the estimate of the Doppler domain steering vector $\hat{\tilde{\mathbf{s}}}_m = [\hat{\tilde{s}}_1 \; \hat{\tilde{s}}_2 \; \dots \; \hat{\tilde{s}}_N]^{\mathrm{T}}$, where $\hat{\tilde{\mathbf{s}}}_m = \mathrm{fftshift}(\mathrm{fft}(\hat{\mathbf{s}}_m))$ with $\hat{\mathbf{s}}_m$ being the time domain steering vector estimate: $\hat{\mathbf{s}}_m = [1 \; e^{j2\pi\hat{F}_m} \; e^{j2\pi 2\hat{F}_m} \; \dots \; e^{j2\pi(N-1)\hat{F}_m}]^{\mathrm{T}}$ is entirely determined by the Doppler frequency estimate $\hat{F}_m$ associated with the local maximum of the $m$-th RPTD.

For the $m$-th RPTD $\mathcal{D}_m$, the DDL-GLR detector of an order $n$ is given by

$$\tilde{\Lambda}_{\mathrm{GLR}m} = \frac{1}{1+\frac{\tilde{\mathbf{y}}_m^{\mathrm{H}} \hat{\tilde{\boldsymbol{\Sigma}}}_m^{-1} \tilde{\mathbf{y}}_m}{K}} \frac{\left|\hat{\tilde{\mathbf{t}}}_m^{\mathrm{H}} \; \hat{\tilde{\boldsymbol{\Phi}}}_m^{-1} \tilde{\mathbf{y}}_m\right|^2}{\hat{\tilde{\mathbf{t}}}_m^{\mathrm{H}} \; \hat{\tilde{\boldsymbol{\Phi}}}_m^{-1} \hat{\tilde{\mathbf{t}}}_m} \underset{\mathrm{H}_0}{\overset{\mathrm{H}_1}{\gtrless}} \tilde{\lambda}_{\mathrm{GLR}} \tag{15}$$

where $\tilde{\lambda}_{\mathrm{GLR}}$ is the detection threshold computed for the given order $n$, the training sample size $K$, and the specified probability of false alarm.

A simple alternative to the GLR detector is the AMF detector. These detectors possess the CFAR property and have comparable detection performance for target signals aligned with the assumed steering vector [2]. However, as shown in [23], the GLR detector provides essential rejection for signals misaligned with the assumed steering vector, whereas the AMF detector does not [2]. Because of measurement errors, the DDL steering vector estimate $\hat{\tilde{\mathbf{t}}}_m$ is not aligned with the actual DDL steering vector $\tilde{\mathbf{t}}_m$ corresponding to the target signal. Therefore, for the detector (15), one can expect essential performance degradation.

The present paper also considers the DDL implementation of the AMF detector. Similarly, for the $m$-th RPTD $\mathcal{D}_m$, the DDL-AMF detector of an order $n$ is given by

$$\tilde{\Lambda}_{\mathrm{AMF}m} = \frac{\left|\hat{\tilde{\mathbf{t}}}_m^{\mathrm{H}} \; \hat{\tilde{\boldsymbol{\Phi}}}_m^{-1} \tilde{\mathbf{y}}_m\right|^2}{\hat{\tilde{\mathbf{t}}}_m^{\mathrm{H}} \; \hat{\tilde{\boldsymbol{\Phi}}}_m^{-1} \hat{\tilde{\mathbf{t}}}_m} \underset{\mathrm{H}_0}{\overset{\mathrm{H}_1}{\gtrless}} \tilde{\lambda}_{\mathrm{AMF}} \tag{16}$$

where $\tilde{\lambda}_{\mathrm{AMF}}$ is the corresponding detection threshold computed for the given order $n$, the training sample size $K$, and the specified probability of false alarm.

Under hypothesis $\mathrm{H}_1$, the DDL CUT vector $\tilde{\mathbf{y}}_m$ associated with the $m$-th RPTD $\mathcal{D}_m$ should contain an essential portion of the total signal power carried by the corresponding time domain CUT vector $\mathbf{x}$ of size $N \times 1$ even when $n \ll N$. Hence, one can expect a minor performance degradation in the DDL-AMF detector relative to its time domain counterpart TD-AMF detector.

For the DDL-AMF detector, calculating the detection threshold $\tilde{\lambda}_{\mathrm{AMF}}$ in (16) from (B9) given in Appendix B entails time-consuming numerical iterations. This approach is not acceptable for a real-time implementation of the DDL-AMF detector. Appendix D provides a computationally simple and accurate approximating formula for computing $\tilde{\lambda}_{\mathrm{AMF}}$.

## VI. PERFORMANCE OF RPTD-BASED DDL DETECTORS

In this section, we analyze the performance of the detectors under investigation: the RPTD-based DDL-GLR detector (15) and the RPTD-based DDL-AMF detector (16). We compare their performance to the optimum detector of order $N$ and the optimum DDL detector of order $n$. This analysis includes known and unknown target Doppler frequency scenarios for the Swerling I target model. Appendices A, B, and C provide equations for calculating detection performance under known target Doppler. We use Monte-Carlo simulations to evaluate detection performance for unknown target Doppler frequency.

The FFT of size $N_{\text{fft}} = 4N$ is used to compute Doppler profiles required to identify local range-Doppler peaks and perform target Doppler measurements. A Matlab-embedded function, "findpeaks," is exploited for identifying local peaks. The fine Doppler estimator from [22] is used for Doppler measurements. In determining the RPTD, we employ the RPTD identification procedure described in Section IV.

We first compare the detection performance of the DDL-GLR detector that employs the RODI concept proposed in [4] with that of the detectors under investigation for the scenario with known target Doppler. Fig. 9 compares the on-grid $P_{\text{D}}$-vs-$F$ plots ($F$ is the normalized Doppler frequency) of the RPTD-based DDL-GLR and the RODI-based DDL-GLR detectors in the presence of strong clutter under the settings identical to that in Fig. 3. In Fig. 9, it can be seen that both detectors exhibit similar performance at frequencies on the DFT grid corresponding to bins in the RODI regions. However, the RPTD-based DDL-GLR detector outperforms the RODI-based DDL-GLR one at the off-grid points. The advantage of the former against the latter is evident in Fig. 10, which compares their $P_{\mathrm{D}}$-vs-$F$ plots for continuous target Doppler frequency (on- and off-grid performance comparison).

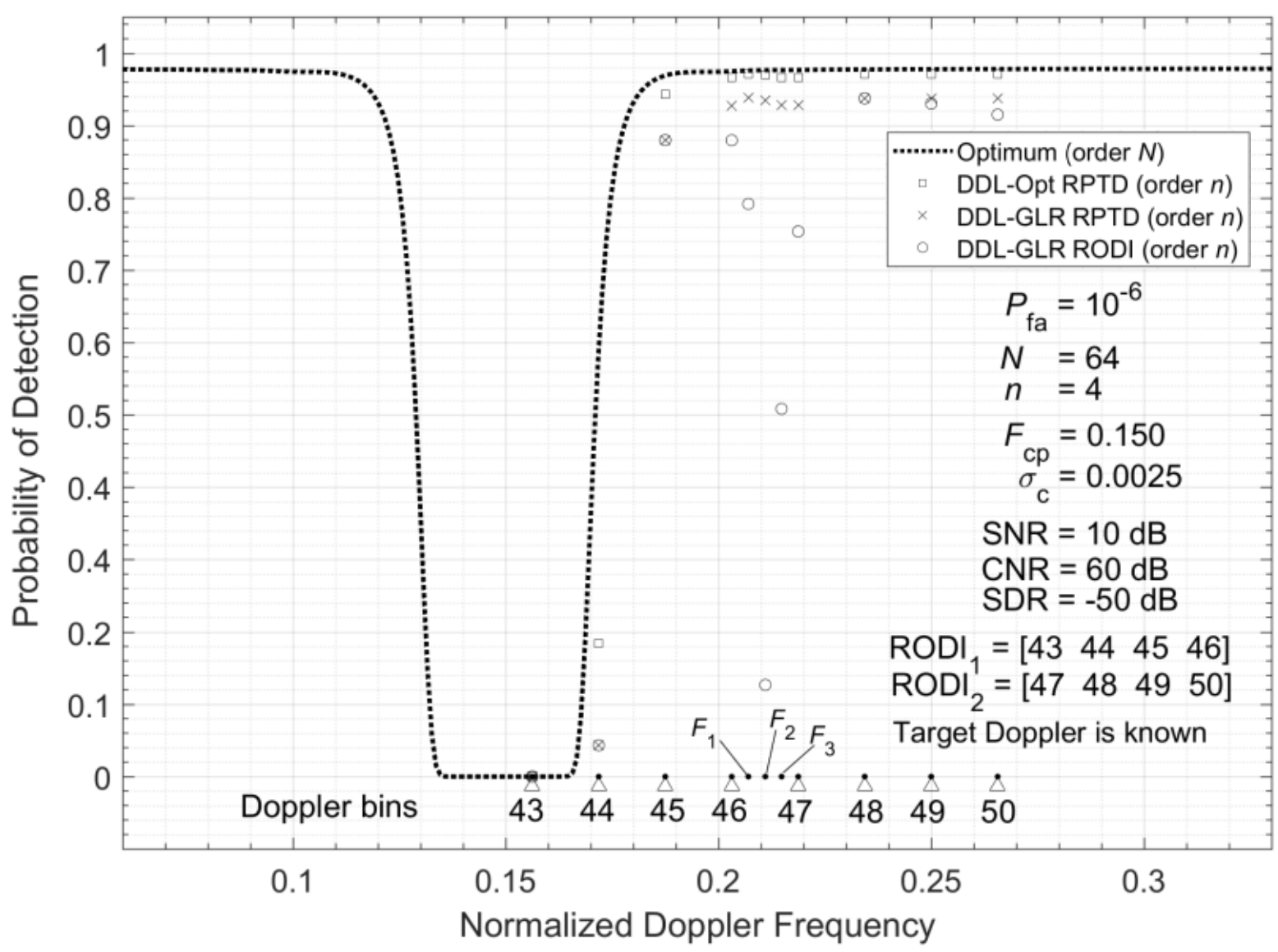

Fig. 9. On-grid $P_{\text{D}}$-vs-$F$ plots comparison of RPTD- and RODI-based DDL detectors

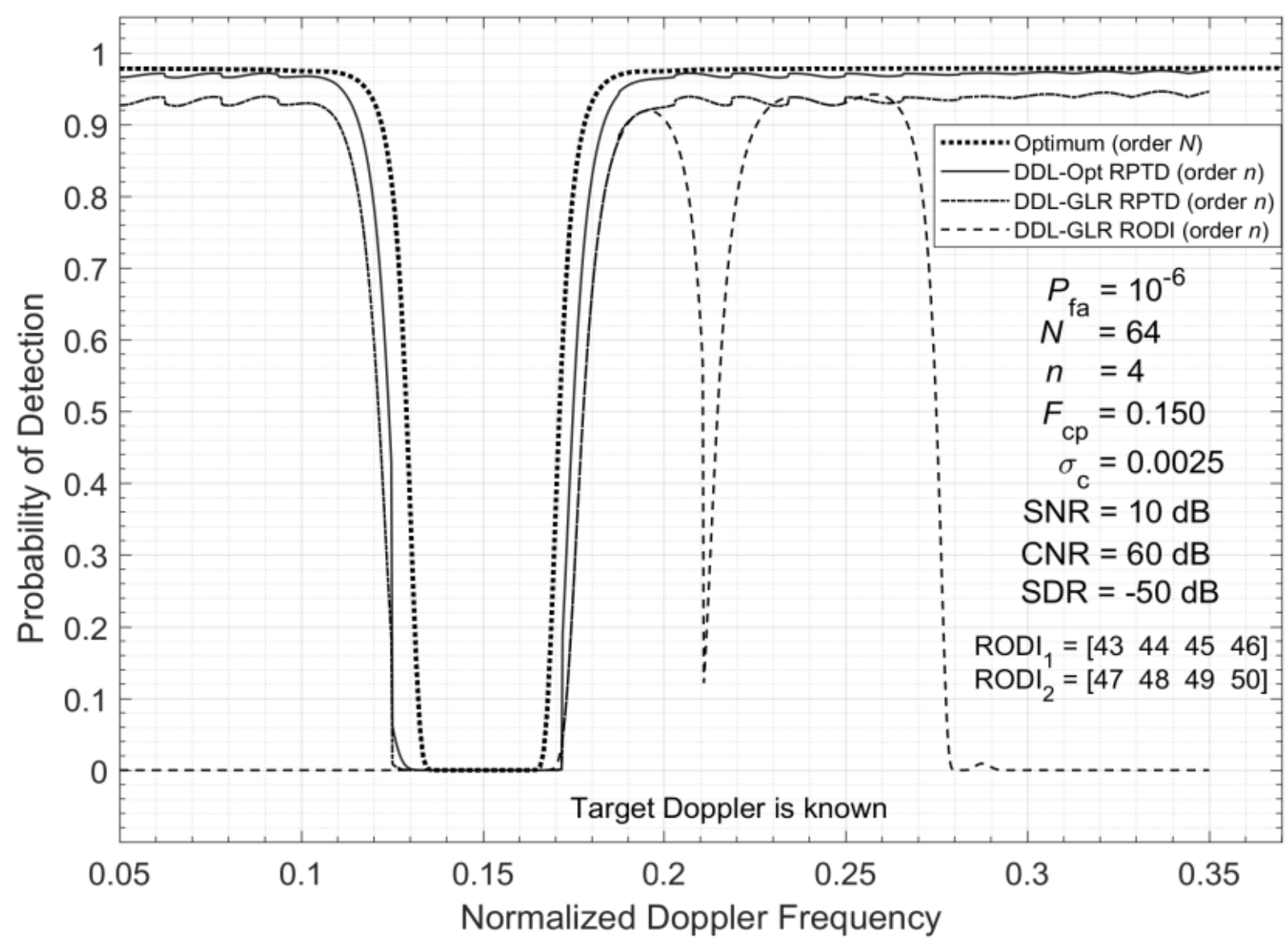

Fig. 10. On- and off-grid $P_{\mathrm{D}}$-vs-$F$ plots comparison of RPTD- and RODI-based DDL detectors

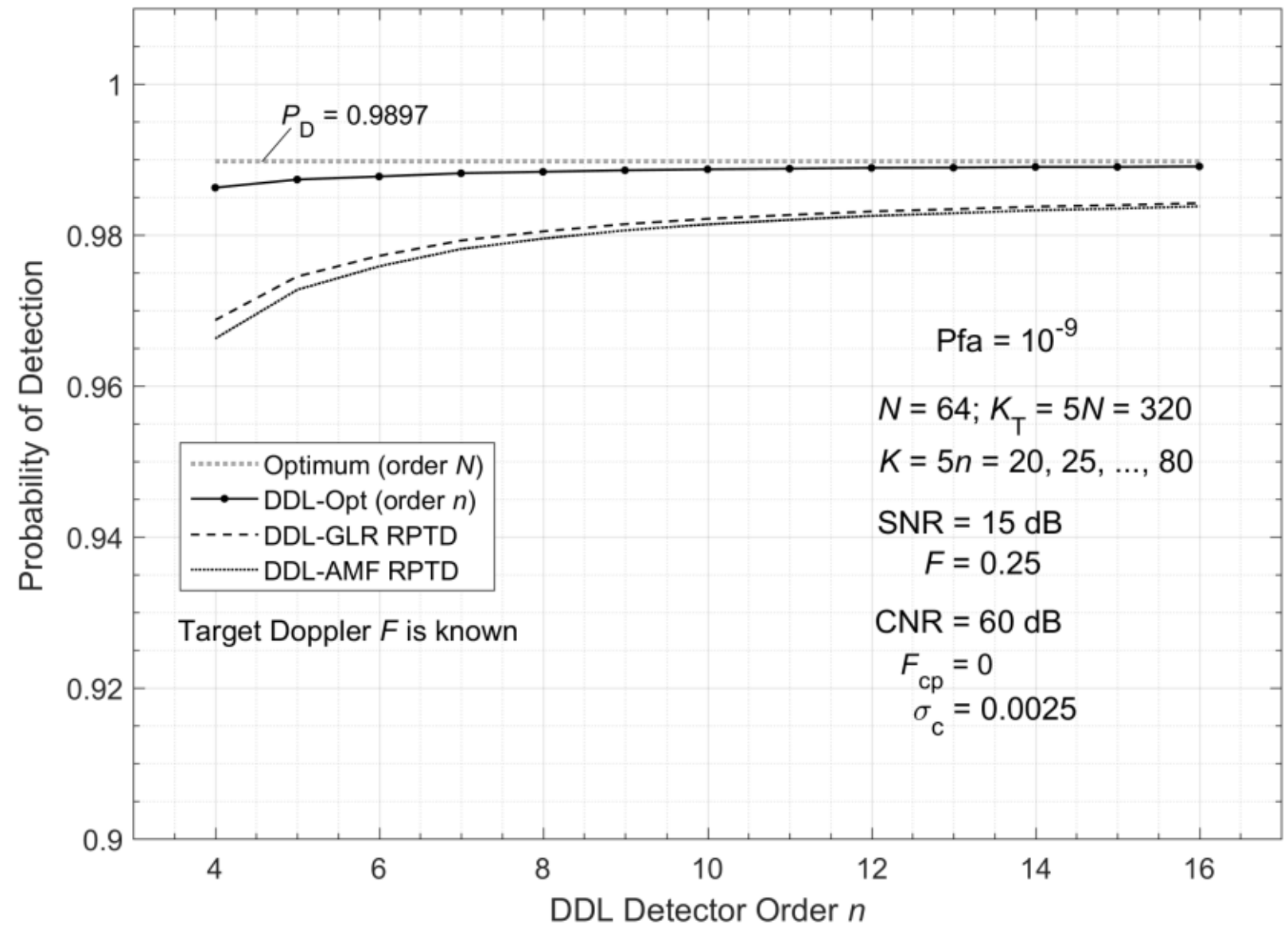

Fig. 11. $P_{\mathrm{D}}$-vs-$n$ plots for RPTD-based DDL-GLR and DDL-AMF detectors (known Doppler frequency)

Fig. 11 shows the $P_{\mathrm{D}}$-vs-$n$ plots ($n$ is the order of a DDL detector) for the probability of false alarm $P_{\mathrm{fa}} = 10^{-9}$. The plots are calculated for the known Doppler frequency scenario ($F = 0.25$) at the signal-to-noise ratio SNR = 15dB (input SDR= -45 dB). The strong clutter (CNR = 60 dB) is assumed to have a zero-centered Gaussian-shaped PSD, which spectral width $\sigma_{\mathrm{c}} = 0.0025$. As can be seen, both of the detectors under investigation exhibit comparable and near-optimum detection performance when the target Doppler frequency is known.

Fig. 12 plots the loss in the detection probability for the detectors under investigation relative to the optimum detector of order $N$ with different choices of $N$. As seen, for $n \geq 3$, the loss in the detection probability does not exceed 5% for $N = 64$ and 128. However, for $n \geq 4$, this loss is below 5% for $N = 32$ and does not exceed 2.5% for $N = 64$ and 128. The probability of detection for the optimum detector of order $N$ is $P_{\mathrm{D}} = 0.9897$.

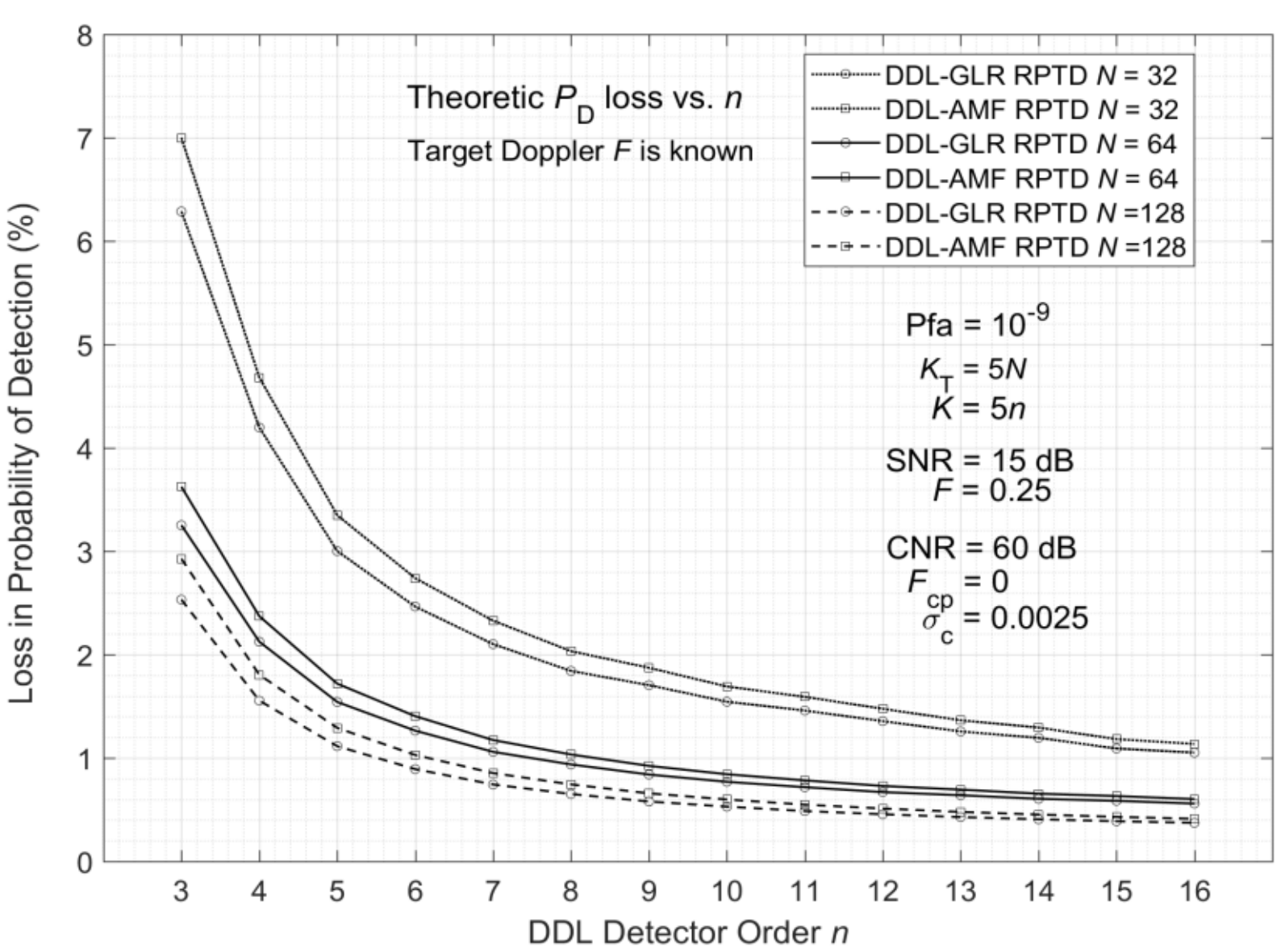

Fig. 12. Loss in $P_{\mathrm{D}}$ for RPTD-based DDL-GLR and DDL-AMF detectors (known Doppler frequency)

Fig. 13 compares the $P_{\mathrm{D}}$-vs-$n$ plots estimated using 10,000 Monte-Carlos for the unknown target Doppler frequency (marked UnKnown) with those for the known target Doppler (marked Known) scenario. As can be seen, the RPTD-based DDL-AMF detector is robust to Doppler measurement errors, while the RPTD-based DDL-GLR one severely degrades.

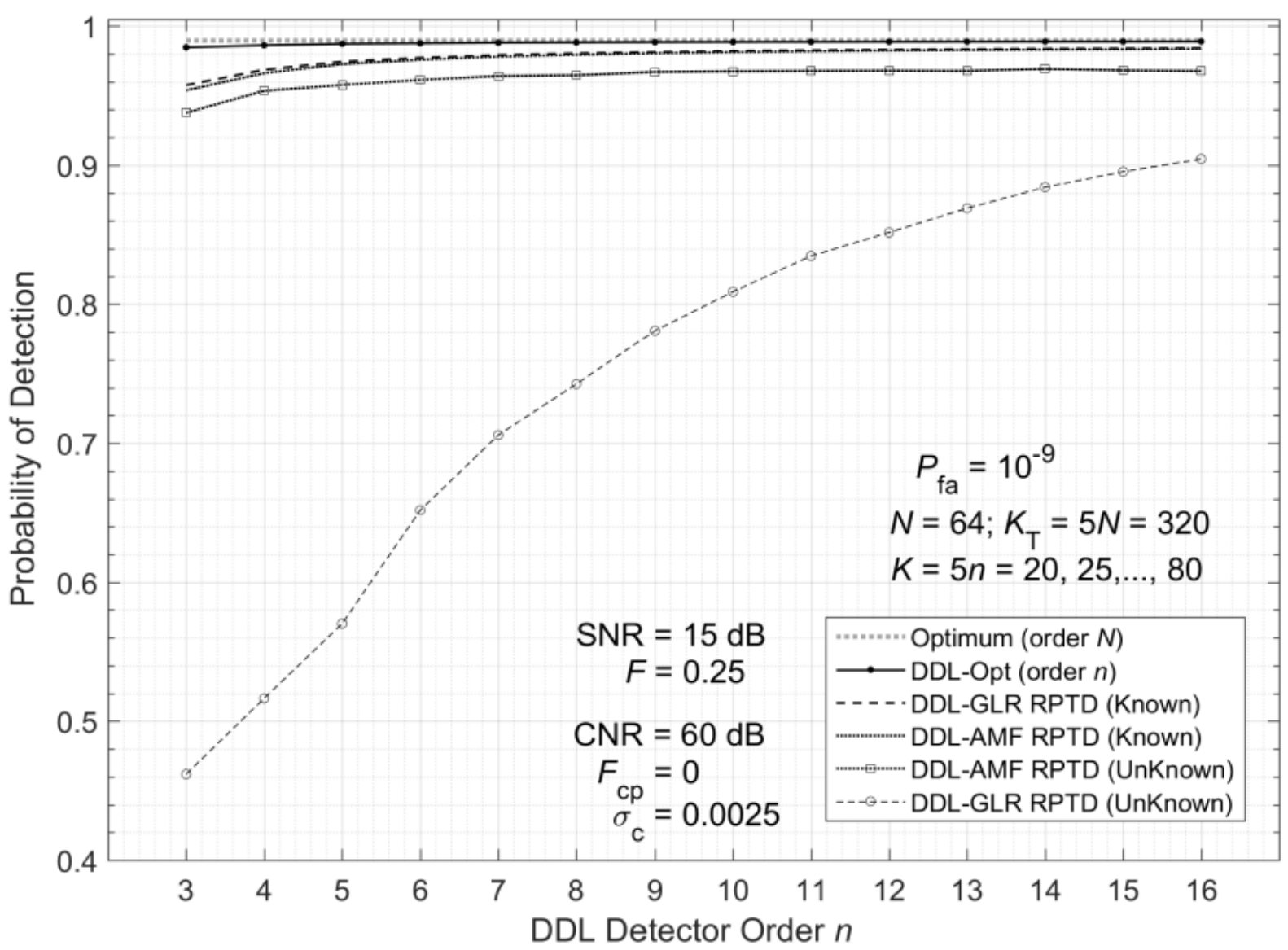

Fig. 13. $P_{\mathrm{D}}$-vs-$n$ plots for RPTD-based DDL-GLR and DDL-AMF detectors

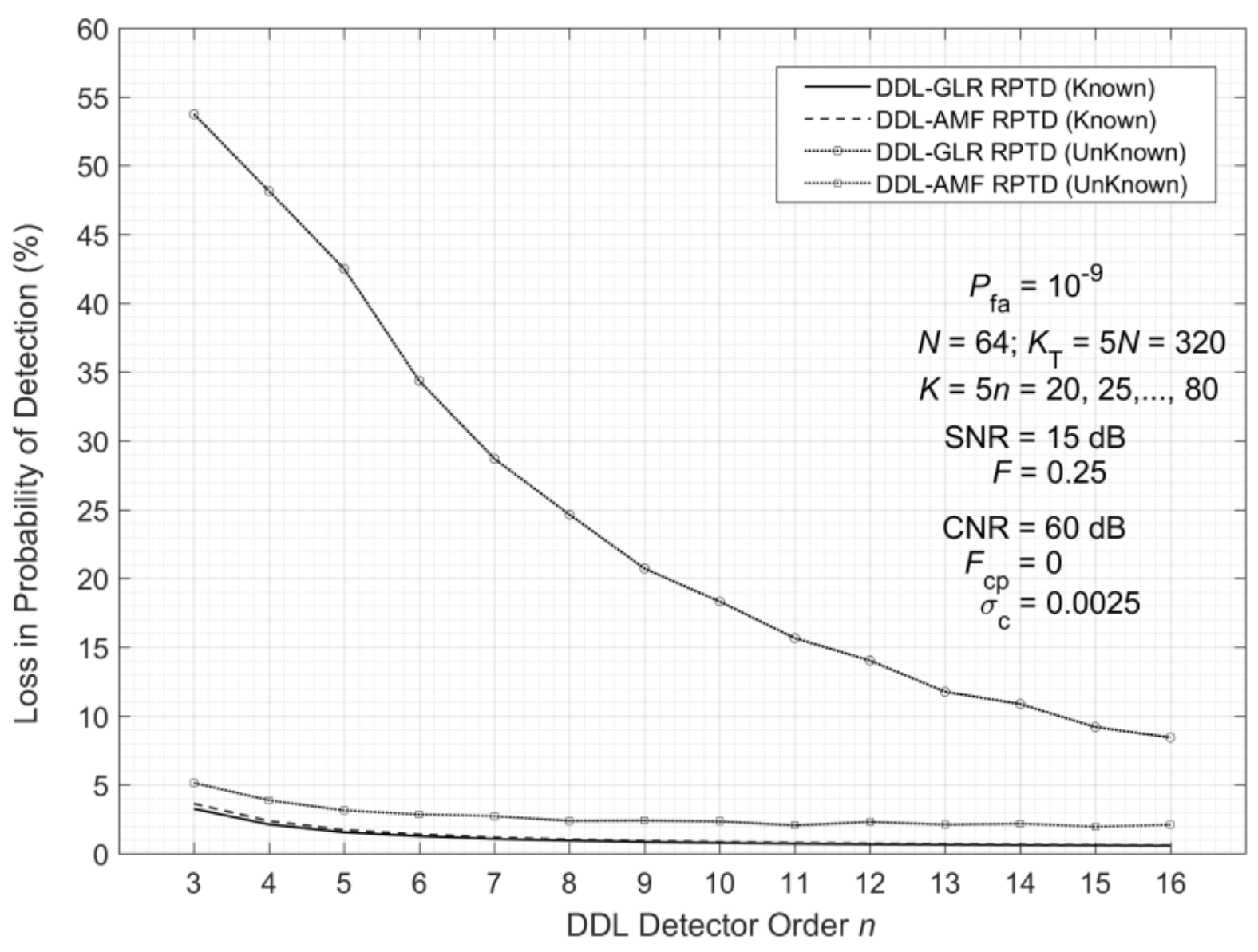

Fig. 14. Loss in $P_{\mathrm{D}}$ for RPTD-based DDL-GLR and DDL-AMF detectors (unknown target Doppler)

Indeed, as seen in Fig. 14, when the target Doppler frequency is unknown, the $P_{\mathrm{D}}$ loss for the DDL-AMF detector does not exceed 5% for $n \geq 4$, while for the DDL-GLR one, it is about 48% for $n = 4$ and 25% even for $n = 8$. The $P_{\mathrm{D}}$ losses are calculated relative to $P_{\mathrm{D}} = 0.9897$ for the optimum detector of order $N$.

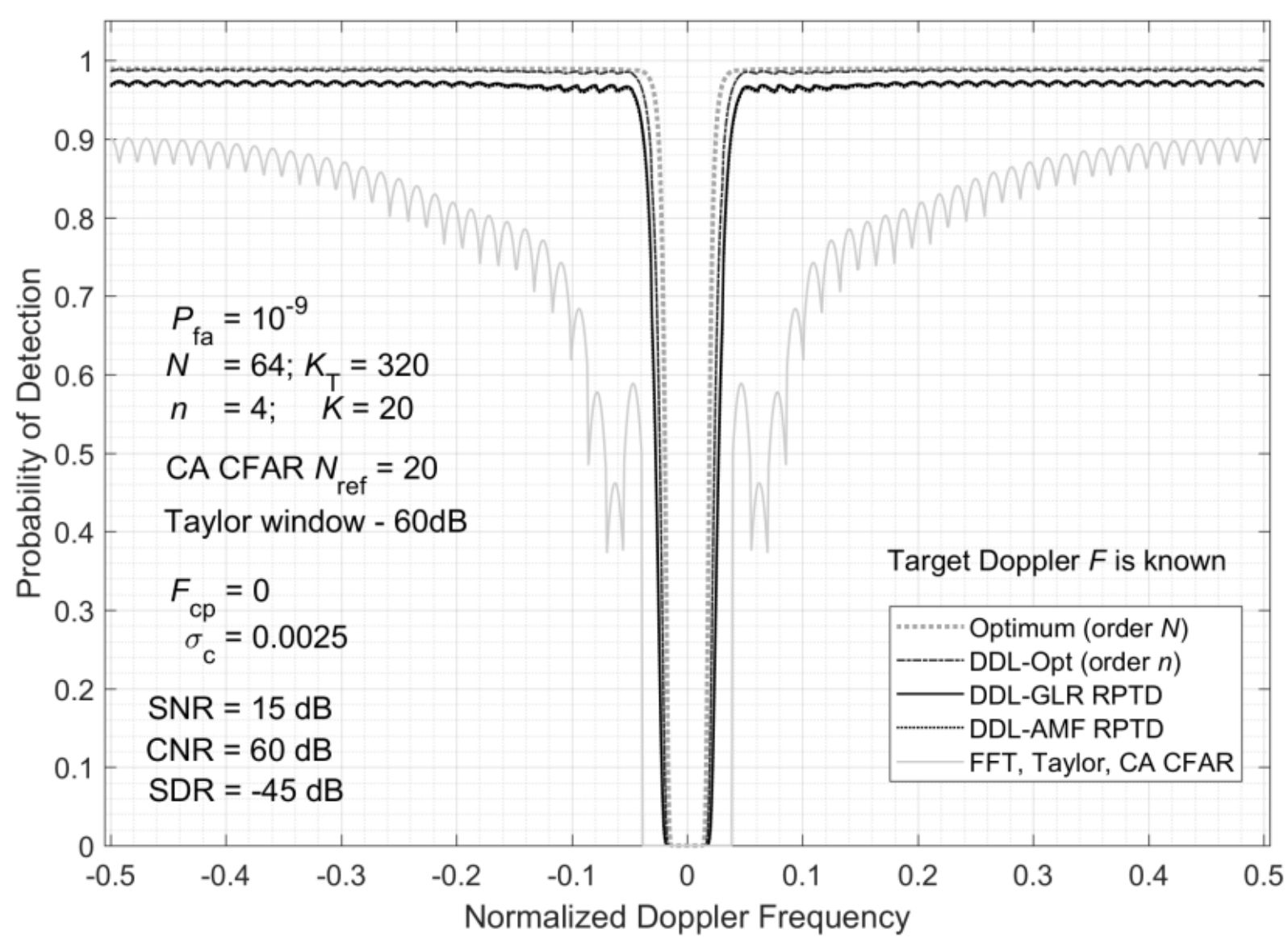

Fig. 15. Comparison of $P_D$-vs-$F$ plots for RPTD-based DDL-GLR and AMF detectors, optimum detectors, and Taylor windowed FFT followed by CA CFAR detector (known target Doppler)

For the detectors under investigation of order $n = 4$, Fig. 15 shows the $P_D$-vs-$F$ plots in the case of the known target Doppler embedded in the strong clutter with a zero-centered Gaussian PSD and spectral width $\sigma_c = 0.0025$. For the Swerling I target, the input signal-to-noise ratio is supposed to be SNR = 15 dB (input SDR = –45 dB). The training sample size is $K_T = 5N = 320$ for the optimum detector of order $N$ and $K = 5n = 20$ for the DDL detectors. As can be seen, both of the detectors under investigation have minor ripples in their frequency responses ($P_D$-vs-$F$ plots). However, they exhibit near-optimum detection performances.

Fig. 15 includes the $P_D$-vs-$F$ plot for a detector employing Taylor windowed $N$-point FFT followed by a CA CFAR that uses $N_{ref} = K = 20$ reference samples. As one can see, this traditional detector does not provide near-optimum performance over the flat spectrum regions associated with the Doppler intervals $-0.5 < F < -0.06$ and $0.06 < F < 0.5$. Moreover, the frequency response of this detector exhibits noticeable ripples due to the inevitable straddling loss. Therefore, the critical assumption [4] that adaptive DDL detection should be applied only to sharply changing spectrum regions is generally incorrect. Fig. 15 demonstrates that the probability of detection for conventional detectors may fall far below the optimum even when the target Doppler is far away from the clutter spectrum peaks. Thus, adaptive DDL detection should generally be applied over the entire Doppler domain.

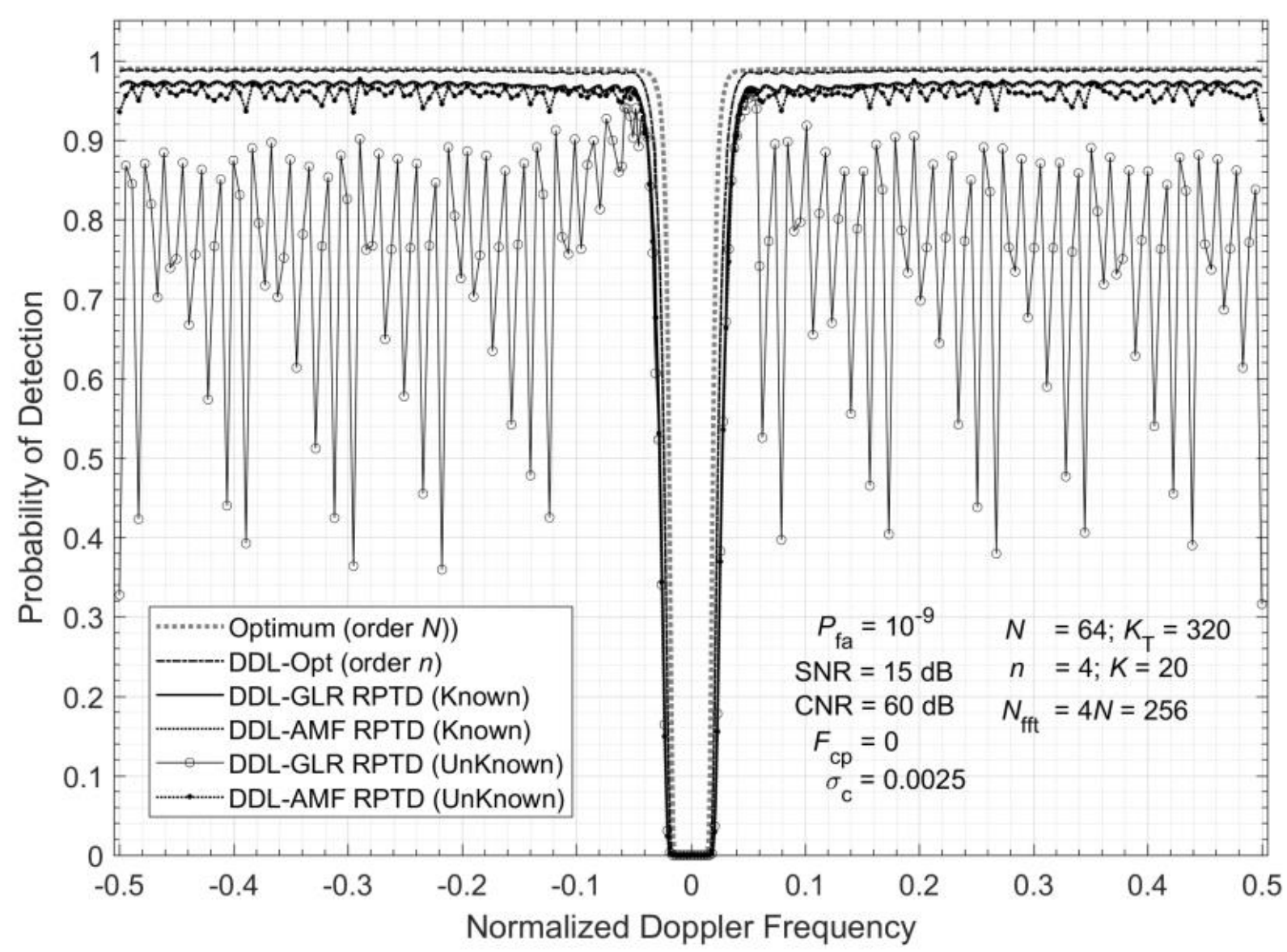

Fig. 16. $P_{\mathrm{D}}$-vs-$F$ plots comparison for known and unknown target Doppler at $N_{\mathrm{fft}} = 4N$

Fig. 16 compares the $P_{\mathrm{D}}$-vs-$F$ plots (estimated using 2,000 Monte-Carlos) for the unknown target Doppler frequency against those for the known target Doppler shown in Fig. 15. As one can see, the DDL-AMF detector's performance for the unknown target Doppler frequency is close to that for the case of known Doppler. The minor ripples in the detection curve do not deprive its near-optimum behavior. However, the DDL-GLR detector's performance is severely impaired. Thus, the former significantly outperforms the latter in detection capability for scenarios with unknown target Doppler.

Fig. 17 compares the $P_{\mathrm{D}}$-vs-$F$ plots of the RPTD-based DDL-AMF detector of order $n = 4$ evaluated for the unknown target Doppler frequency (using 2,000 Monte Carlos) at two different lengths $N_{\mathrm{fft}} = 4N$ and $2N$ of the FFT for fine Doppler measurements. As expected, better detection performance is achieved for the former.

Fig. 18 demonstrates that the RPTD-based DDL-AMF detector of order $n = 4$ can maintain reliable detection performance in three clutters even for the unknown target Doppler frequency. In this multiple-clutter scenario, the RPTD-based DDL-GLR detector also degrades when the target Doppler frequency is unknown, as in situations with a single clutter.

For scenarios with the known and unknown target Doppler frequency $F$, Fig. 19 (a) and (b) show the $P_{\mathrm{D}}$-vs-SDR plots (SDR stands for the signal-to-disturbance ratio) evaluated for $F = 0.15$ and $F = 0.17$, respectively. These plots confirm the significant advantage of the DDL-AMF detector against the DDL-GLR one under the unknown target Doppler frequency.

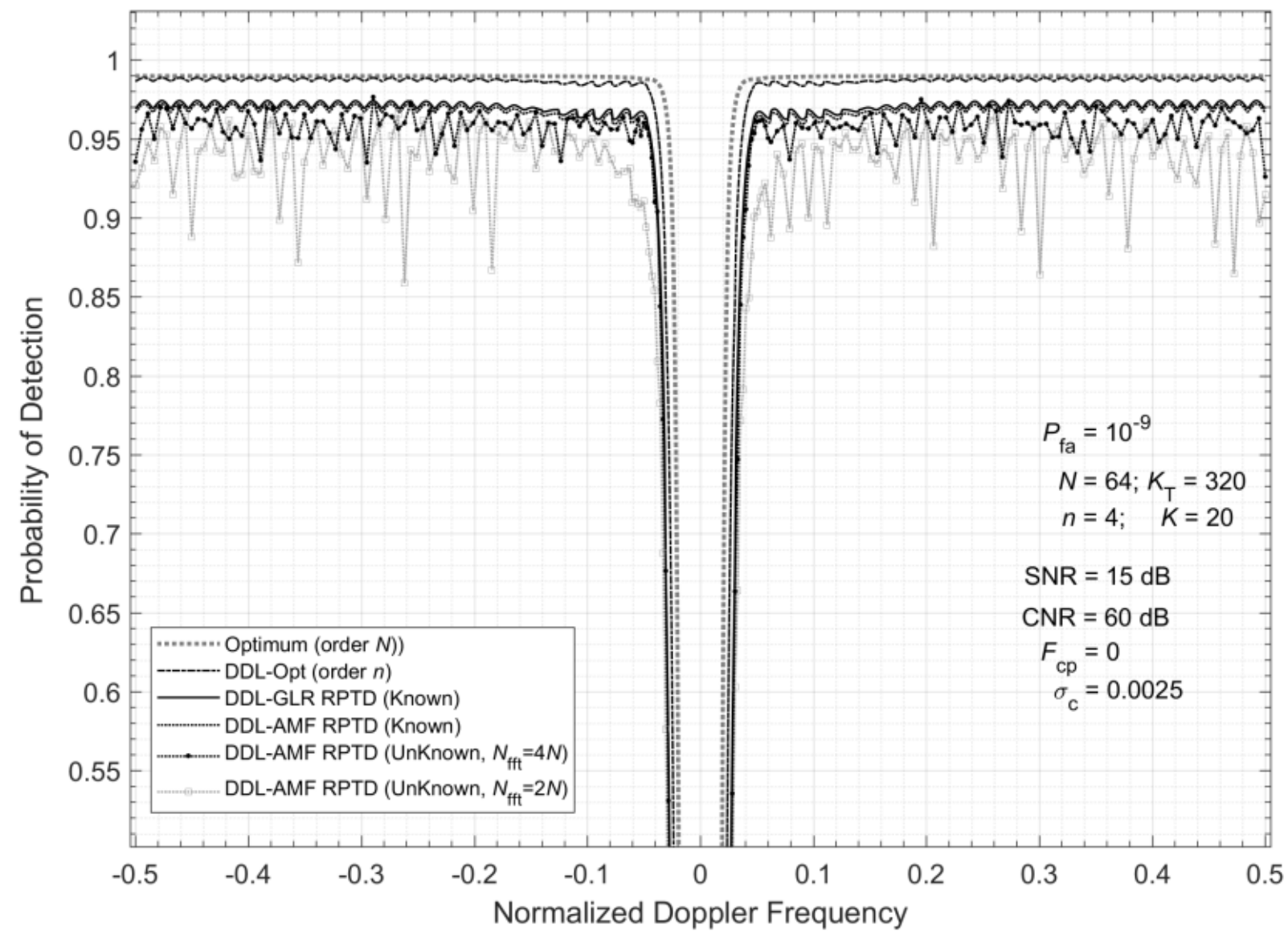

Fig. 17. $P_{\mathrm{D}}$-vs-$F$ plots comparison at different $N_{\mathrm{fft}}$ for fine Doppler measurements

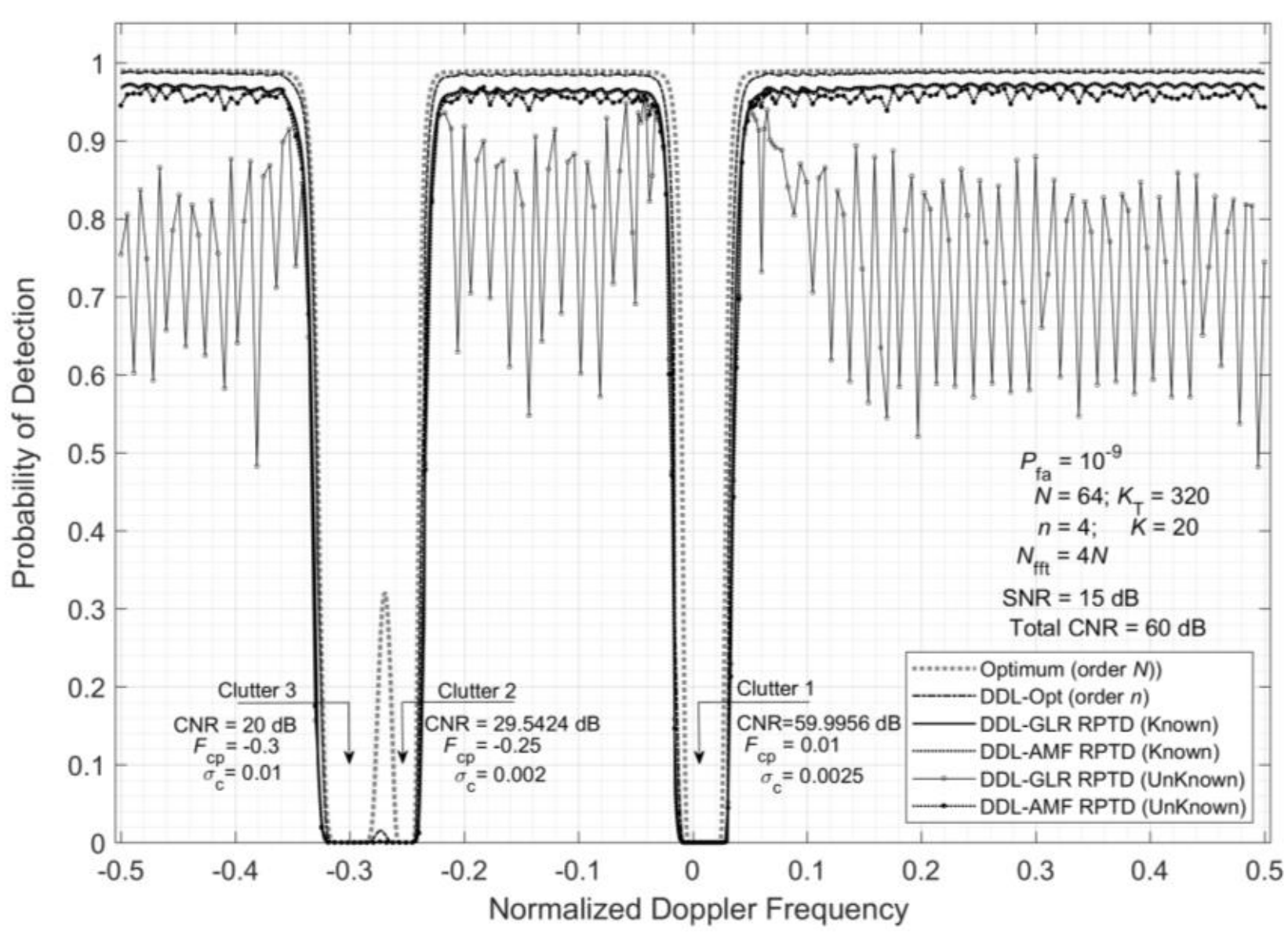

Fig. 18. $P_{\mathrm{D}}$-vs-$F$ plots for detectors under investigation in multiple-clutter scenario

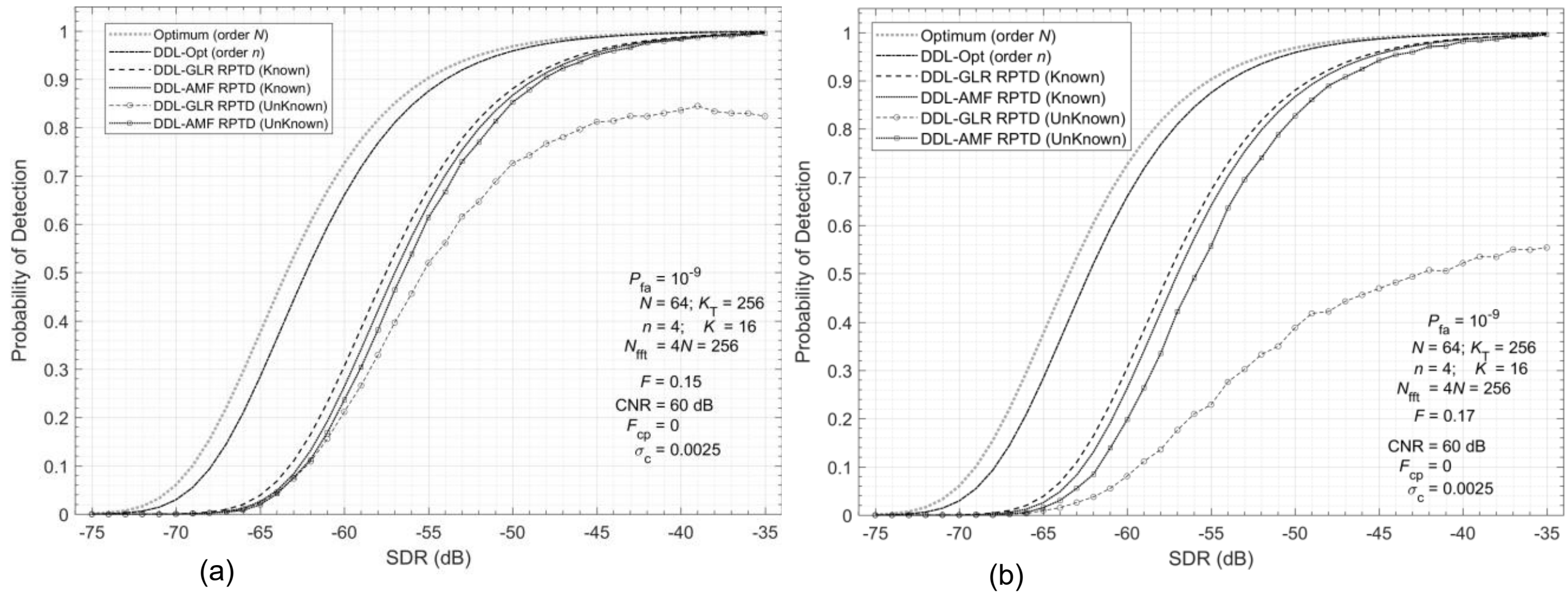

Fig. 19. $P_{\mathrm{D}}$-vs-SDR plots for detectors under investigation at different target Doppler

## VII. COMPUTATIONAL LOAD

This Section evaluates and compares the computational load of the conventional TD-AMF detector and the RPTD-based DDL-AMF detector proposed in this paper. This evaluation considers the total amount of dominant calculations per CPI. For the TD-AMF detector, the CPI data are represented by the fast/slow time (range-pulse) matrix $\mathbf{X}$ of size $M\times N$. For the DDL-AMF detector, the CPI data are represented by the $M\times N$ Range-Doppler matrix $\widetilde{\mathbf{Y}}$, which results from the row-wise FFT applied to the matrix $\mathbf{X}$. We assume that the number of representative range cells $M_{\mathrm{R}}$ and the average number of local Doppler peaks per each representative range cell $N_{\mathrm{D}}$ are known. Hence, the number of local range-Doppler peaks per CPI is $N_{\mathrm{p}} = N_{\mathrm{D}}M_{\mathrm{R}}$.

For the TD-AMF detector implementation, the $N\times N$ sample covariance matrix (SCM) must be calculated, which requires about $K_{\mathrm{T}}N^2$ complex-valued floating-point operations (flops); hence, $K_{\mathrm{T}} = 4N$ incurs $4N^3$ flops. Computing the inverse of the SCM requires about $2N^3$ flops. Thus, calculating the SCM and the inverse matrix involve about $6N^3$ flops per cell under test (CUT). Since the TD-AMF operates on data in the matrix $\mathbf{X}$, the CUT vectors for this detector are those rows in $\mathbf{X}$ whose indices are in the set of representative range cell indices $\mathcal{R} = \{i_k, k = 1,2,\ldots,M_{\mathrm{R}}\}$. Hence, these operations involve $\mathrm{CL1_{TD}} = 6N^3M_{\mathrm{R}}$ flops per CPI. Because the target Doppler frequency is unknown, the TD-AMF must calculate its test statistic $N_{\mathrm{D}}$ times (on average) per CUT or exactly $N_{\mathrm{p}}$ times per CPI. These calculations require $\mathrm{CL2_{TD}} = 2N^2N_{\mathrm{p}}$ flops per CPI. Calculating each test statistic requires the $N\times 1$ steering vector estimate associated with the corresponding local range-Doppler peak. In turn, this estimate uses the corresponding Doppler frequency estimate. The fine Doppler estimator needs data only from one $N_{\mathrm{fft}}$-point FFT to obtain these Doppler frequency estimates ($N_{\mathrm{D}}$ times on average) for each representative range cell. One such FFT incurs $(3/2)N_{\mathrm{fft}} \log_2 N_{\mathrm{fft}}$ flops. Hence, estimating the steering vectors incurs $\mathrm{CL3_{TD}} = M_R \times (3/2)N_{\mathrm{fft}} \log_2 N_{\mathrm{fft}}$ flops per CPI. Thus, the total dominant computation load per CPI for the TD-AMF detector is

$$\mathrm{CL_{TD}} = \mathrm{CL1_{TD}} + \mathrm{CL2_{TD}} + \mathrm{CL3_{TD}} \tag{17}$$

For the RPTD-based DDL-AMF detector, calculating the corresponding SCM of size $n\times n$ requires about $Kn^2$ flops; hence, $K = 4n$ incurs $4n^3$ flops. Computing the inverse matrix involves about $2n^3$ flops. Calculating DDL test statistics requires $2n^2$ flops per CUT. Therefore, calculating the SCM, the inverse matrix and test statistics involves about $6n^3 + 2n^2$ flops per CUT or $\mathrm{CL1_{DDL}} = N_{\mathrm{p}} \times (6n^3 + 2n^2)$ flops per CPI because each DDL CUT is associated with a local range-Doppler peak.

To evaluate the dominant computational load for the DDL-AMF detector, we must also consider the FFT operations involved in its implementation. First, these are about $\mathrm{CL2_{DDL}}= M \times ((3/2)N \log_2 N)$ flops per CPI for the $N$-point FFT that calculates the $M\times N$ detection matrix $\widetilde{\mathbf{Y}}$. Second, the fine Doppler estimator employs the $N_{\mathrm{fft}}$-point FFT that incurs additional $(3/2)N_{\mathrm{fft}} \log_2 N_{\mathrm{fft}}$ flops per representative range cell or $\mathrm{CL3_{DDL}}= M_{\mathrm{R}} \times (3/2)N_{\mathrm{fft}} \log_2 N_{\mathrm{fft}}$ flops per CPI. Third, calculating the DFT images for the estimated target steering vectors incurs $\mathrm{CL4_{DDL}}=N_{\mathrm{p}} \times ((3/2)N \log_2 N)$ flops per CPI. Thus, the total dominant computation load per CPI for the DDL-GLR detector is

$$\mathrm{CL_{DDL}} = \mathrm{CL1_{DDL}} + \mathrm{CL2_{DDL}} + \mathrm{CL3_{DDL}} + \mathrm{CL4_{DDL}} \tag{18}$$

Using (17) and (18), we calculate the DDL gain in the computational load as

$$G_{\mathrm{DDL}} = \mathrm{CL_{TD}}/\mathrm{CL_{DDL}} \tag{19}$$

Before evaluating the order of magnitude of achievable DDL gain, we note that the upper bound for the number of representative range cells per CPI is $\sup M_R = M$. Therefore, assuming that an actual $M_R$ value makes up a definite $\gamma$ percent of its maximum is reasonable. Thus, we get $M_{\mathrm{R}} = \mathrm{round}\left(\frac{\gamma}{100}M\right)$. We evaluate the DDL gain in computational load using (19) for $n = 4, 5$ and 6, at $N \in \{64, 128, 256\}$, $N_{\mathrm{fft}} = 4N$, $M$=8,000, $\gamma = 90\%$, and $N_{\mathrm{D}} = N$. This setting for $N_{\mathrm{D}}$ is reasonable for $N_{\mathrm{fft}} \geq 2N$.

TABLE I
DDL GAIN IN COMPUTATIONAL LOAD

| $n$ | $N$ | | |
|---|---|---|---|
| | 64 | 128 | 256 |
| 4 | 31 | 71 | 147 |
| 5 | 22 | 59 | 132 |
| 6 | 16 | 47 | 116 |

Table I summarizes the $G_{DDL}$ values rounded to the nearest smallest integer. As can be seen, the RPTD-based DDL-AMF detector can provide faster adaptive detection than its time domain counterpart due to a much lower computational load. This is another advantage that makes the former the preferable detector for actual implementation in radar systems.

## VIII. CONCLUSION

This paper has proposed a new Doppler domain localized (DDL) adaptive implementation of the classical AMF detector. The proposed DDL-AMF detector employs the concept of a region of possible target detection (RPTD), a small set of Doppler bins that capture most of the target signal power. This new approach to adaptive detection ensures near-optimum detection of targets in various scenarios, including training data-deficient scenarios, clutter environments with multimodal power spectral density, and unknown Doppler frequencies of targets. The computational load required to implement the proposed DDL-AMF detector is essentially lower than that of the classical AMF detector.

The proposed rapid adaptive DDL-AMF detector offers a promising solution for detecting radar targets embedded in strong non-homogeneous clutter.

Future work may extend the RPTD-based adaptive DDL processing principle to target detection under interferences governed by complex elliptically symmetric distribution. This approach can be developed for adaptive target detection in compound-Gaussian clutter and extended to spatial-temporal adaptive processing for airborne surveillance radar systems.

**Appendix A**

Optimum Detectors

A1. Time-Domain Optimum Detector of Order $N$

From the Neyman-Pearson lemma, the optimum strategy for detecting a coherent signal embedded in Gaussian disturbance is based on the whitening-matched filter, also known as the filter maximizing the output signal-to-disturbance ratio (SDR). For a given disturbance covariance matrix $\mathbf{\Sigma}$, $N\times N$ with disturbance being Gaussian clutter plus thermal noise, the $N\times 1$ weight vector of the optimum filter is calculated as [24]

$$\mathbf{w} = \kappa\mathbf{\Sigma}^{-1}\mathbf{s} \tag{A1}$$

where $\kappa$ is a scalar that does not affect the output SDR, $\mathbf{s} = [1 \;\; e^{j2\pi F} \;\; e^{j2\pi 2F} \dots e^{j2\pi(N-1)F}]^{\mathrm{T}}$ is the $N\times 1$ the desired (or steering) signal vector for a known Doppler frequency $F$ (the superscript $^{\mathrm{T}}$ stands for the matrix transposition).

We specify the disturbance covariance matrix $\mathbf{\Sigma}$ as

$$\mathbf{\Sigma} = P_{\mathrm{c}}\mathbf{C}_0 + P_{\mathrm{n}}\mathbf{I} \tag{A2}$$

where $P_{\mathrm{c}}$ and $P_{\mathrm{n}}$ represent the clutter and thermal noise power, respectively, $\mathbf{I}$ is an $N\times N$ identity matrix, and $\mathbf{C}_0$ is an $N\times N$ normalized clutter covariance matrix defined by

$$\mathbf{C}_0 = [c_{mn}] = \left[e^{2(\pi\sigma_c(m-n))^2 + i(m-n)2\pi F_{\mathrm{cp}}}\right] \tag{A3}$$

In (A3), the normalized Doppler frequency $F_{\mathrm{cp}}$, $|F_{\mathrm{cp}}| < 0.5$, defines the center of the clutter spectrum, and $\sigma_c$ is the parameter controlling the clutter spectrum bandwidth.

The output SDR of the optimum filter is given by [24]

$$\gamma_{\mathrm{o}} = P_{\mathrm{s}} \cdot \mathbf{s}^{\mathrm{H}}\mathbf{\Sigma}^{-1}\mathbf{s} \tag{A4}$$

where $P_{\mathrm{s}}$ is the average signal power. Using the normalized disturbance covariance matrix $\mathbf{\Sigma}_0 = \mathbf{\Sigma}/(P_{\mathrm{c}} + P_{\mathrm{n}})$, for which $[\mathbf{\Sigma}_0]_{kk} = 1, k = 1,2,\dots,N$, yields

$$\gamma_{\mathrm{o}} = \gamma_{\mathrm{in}}(\mathbf{s}^{\mathbf{H}}\mathbf{\Sigma}_{\mathbf{0}}^{-\mathbf{1}}\mathbf{s}) \tag{A5}$$

where $\gamma_{\mathrm{in}} = P_{\mathrm{s}}/(P_{\mathrm{c}} + P_{\mathrm{n}})$ is the input SDR. Using the matrix $\mathbf{\Sigma}_0$ allows the following presentation for (A3)

$$\mathbf{w} = \kappa\mathbf{\Sigma}_0^{-\mathbf{1}}\mathbf{s} \tag{A6}$$

To test the received data vector $\mathbf{x}$ for the target presence, the optimum detector first calculates the optimum filter output $Y = \mathbf{w}^{\mathrm{H}}\mathbf{x} = \kappa\mathbf{s}^{\mathrm{H}}\mathbf{\Sigma}_0^{-\mathbf{1}}\mathbf{x}$ and the module-squared value $|Y|^2 = \kappa^2\left|\mathbf{s}^{\mathrm{H}}\mathbf{\Sigma}_0^{-\mathbf{1}}\mathbf{x}\right|^2$.

Then, it compares $|Y|^2$ against the detection threshold $\eta$

$$|Y|^2 \underset{\mathrm{H}_0}{\overset{\mathrm{H}_1}{\gtrless}} \eta \tag{A7}$$

and declares that the hypothesis $\mathrm{H}_1$ (target present) is true if $|Y|^2 \geq \eta$, otherwise it declares that the hypothesis $\mathrm{H}_0$ (no target) is true.

For the optimum detector of order $N$ given by (A7), the probability of detection $P_{\mathrm{D}}$ in the case of the Swerling I target is calculated as [2]

$$P_{\mathrm{D}} = P_{\mathrm{FA}}^{\frac{1}{1+\gamma_0}} \tag{A8}$$

where $P_{\mathrm{FA}}$ is the required false alarm probability.

A2. Optimum DDL Detector

Gaussian distributions are closed under the DFT. Hence, the optimum DDL detector theory can be built by analogy with the time-domain optimum detector. To design an optimum DDL detector, we first define the DFT images for the received time domain data vector $\mathbf{x} = [x_1\ x_2\ ...\ x_N]^{\mathrm{T}}$, the steering vector $\mathbf{s} = [1\ e^{j2\pi F}\ ...\ e^{j2\pi(N-1)F}]^{\mathrm{T}}$, and the disturbance covariance matrix $\mathbf{\Sigma}$ of size $N \times N$, assuming that the target Doppler is known.

Because the present paper uses zero-Doppler-centered DFT data, the DFT image $\tilde{\mathbf{y}}$ associated with the data vector $\mathbf{x}$ is $\tilde{\mathbf{y}} = [\tilde{y}_1\ \tilde{y}_2\ ...\ \tilde{y}_N]^{\mathrm{T}} = \mathrm{fftshift}(\mathrm{fft}(\mathbf{x}))$. Similarly, the DFT image $\tilde{\mathbf{s}}$ associated with the steering vector $\mathbf{s}$ is given by $\tilde{\mathbf{s}} = [\tilde{s}_1\ \tilde{s}_2\ ...\ \tilde{s}_N]^{\mathrm{T}} = \mathrm{fftshift}(\mathrm{fft}(\mathbf{s}))$. Accordingly, for the matrix $\mathbf{\Sigma}$, we have $\tilde{\mathbf{\Sigma}} = \mathrm{fftshift}(\mathrm{fft}(\mathbf{\Sigma}))$. Using formula $\mathbf{\Sigma} = (P_{\mathrm{c}} + P_{\mathrm{n}})\mathbf{\Sigma}_0$ yields the DFT image associated with the normalized disturbance covariance matrix $\mathbf{\Sigma}_0$ as $\tilde{\mathbf{\Sigma}}_0 = \mathrm{fftshift}(\mathrm{fft}(\mathbf{\Sigma}_0)) = \tilde{\mathbf{\Sigma}}/(P_{\mathrm{c}} + P_{\mathrm{n}})$.

Since the target Doppler frequency $F$ is known, one can easily define the RPTD region corresponding to the $m$-th local range-Doppler peak using the identification procedure given in Section IV. This $m$-th RPTD region is represented by an associated set of Doppler bins $\mathcal{D} = \{d_1\ d_2\ ...\ d_n\}$.

Having defined the $m$-th RPTD set $\mathcal{D} = \{d_1\ d_2\ ...\ d_n\}$, it is straightforward to obtain the $n\times 1$ DDL steering vector $\tilde{\mathbf{t}}_m = [\tilde{t}_{m1}\ \tilde{t}_{m2}\ ...\ \tilde{t}_{mn}]^{\mathrm{T}}$ associated with this RPTD by extracting from the $N\times 1$ vector $\tilde{\mathbf{s}} = [\tilde{s}_1\ \tilde{s}_2\ ...\ \tilde{s}_N]^{\mathrm{T}}$ the entries located at the $d_k$-th positions, i.e., $\tilde{t}_{mk} = \tilde{s}_{d_k}$, $k = 1,2,\dots,n$. Similarly, we obtain the $n\times 1$ DDL vector of received data $\tilde{\mathbf{y}}_m = [\tilde{y}_{m1}\ \tilde{y}_{m2}\ ...\ \tilde{y}_{mn}]^{\mathrm{T}}$ by extracting from the $N\times 1$ vector $\tilde{\mathbf{y}} = [\tilde{y}_1\ \tilde{y}_2\ ...\ \tilde{y}_N]^{\mathrm{T}}$ the entries located at the $d_k$-th positions, i.e., $\tilde{y}_{mk} = \tilde{y}_{d_k}$, $k = 1,2,\dots,n$.

The $n\times n$ disturbance covariance matrix $\tilde{\mathbf{\Phi}}_m$ associated with this RPTD comprises the entries located at the intersection of the $d_k$-th rows and $d_l$-th columns, $k, l = 1,2, \dots, n$ in the matrix $\widetilde{\mathbf{\Sigma}} = [\tilde{\sigma}_{mn}]$, $m, n = 1,2, \dots, N$. Thus, the matrix $\widetilde{\mathbf{\Phi}}_m$ associated with the set $\mathcal{D} = \{d_1\ d_2 \dots d_n\}$ is generated by simply extracting the corresponding entries in the matrix $\widetilde{\mathbf{\Sigma}}$ as given below

$$\widetilde{\mathbf{\Phi}}_m = [\tilde{\varphi}_{kl}],\ \tilde{\varphi}_{kl} = \tilde{\sigma}_{d_k d_l},\ k, l = 1,2, \dots, n \tag{A9}$$

Similarly, the $n{\times}n$ DDL matrix $\widetilde{\mathbf{\Phi}}_{0m}$ associated with the $N{\times}N$ matrix $\widetilde{\mathbf{\Sigma}}_0$ is generated as

$$\widetilde{\mathbf{\Phi}}_{0m} = [\tilde{\varphi}^0_{kl}],\ \tilde{\varphi}^0_{kl} = \tilde{\sigma}^0_{d_k d_l},\ k, l = 1,2, \dots, n \tag{A10}$$

where the matrix $\widetilde{\mathbf{\Sigma}}_0 = [\tilde{\sigma}^0_{ij}]$, $i, j = 1,2, \dots, N$ is the DFT image corresponding to the time-domain normalized disturbance covariance matrix $\mathbf{\Sigma}_0$ through the equation $\widetilde{\mathbf{\Sigma}}_0 = \widetilde{\mathbf{\Sigma}}/(P_\mathrm{c} + P_\mathrm{n})$.

Following the theory of the time-domain optimal detector yields that for the $m$-th RPTD, the $n{\times}1$ vector of optimum weights (optimum DDL filter) maximizing the RPTD-associated SDR is given by

$$\widetilde{\mathbf{w}}_m = \kappa \widetilde{\mathbf{\Phi}}_m^{-1} \tilde{\mathbf{t}}_m \tag{A11}$$

where $\widetilde{\mathbf{\Phi}}_m$ is the $n{\times}n$ disturbance covariance matrix and $\tilde{\mathbf{t}}_m$ is the $n{\times}1$ DDL steering vector that both are associated with $m$-th RPTD, and $\kappa$ is an arbitrary scalar that does not affect the output SDR.

The SDR at the output of the optimum DDL filter of order $n$ is given by

$$\tilde{\gamma}_\mathrm{o} = P_\mathrm{s} \cdot \tilde{\mathbf{t}}_m^\mathrm{H} \widetilde{\mathbf{\Phi}}_m^{-1} \tilde{\mathbf{t}}_m \tag{A12}$$

where $P_\mathrm{s}$ is the average signal power (the superscript $^\mathrm{H}$ stands for the Hermitian transposition). Using the matrix $\widetilde{\mathbf{\Phi}}_{0m}$, yields

$$\tilde{\gamma}_\mathrm{o} = \gamma_\mathrm{in} (\tilde{\mathbf{t}}_m^\mathrm{H} \widetilde{\mathbf{\Phi}}_{0m}^{-1} \tilde{\mathbf{t}}_m) \tag{A13}$$

where $\gamma_\mathrm{in} = P_\mathrm{s}/(P_\mathrm{c} + P_\mathrm{n})$ is the input SDR. Using the matrix $\widetilde{\mathbf{\Phi}}_{0m}$ we get for the optimum DDL vector (A13)

$$\widetilde{\mathbf{w}}_m = \kappa \widetilde{\mathbf{\Phi}}_{0m}^{-1} \tilde{\mathbf{t}}_m \tag{A14}$$

To test the DDL vector of the received data $\tilde{\mathbf{y}}_m$ for the target presence, the optimum DDL detector first calculates the optimum DDL filter output $\tilde{Y}_m = \widetilde{\mathbf{w}}_m^\mathrm{H} \tilde{\mathbf{y}}_m = \kappa \tilde{\mathbf{t}}_m^\mathrm{H} \widetilde{\mathbf{\Phi}}_{0m}^{-1} \tilde{\mathbf{y}}_m$ and the module-squared value $|\tilde{Y}_m|^2$. Comparing it against the detection threshold $\tilde{\eta}$

$$|\tilde{Y}_m|^2 \underset{\mathrm{H}_0}{\overset{\mathrm{H}_1}{\gtrless}} \tilde{\eta} \tag{A15}$$

results in declaring the hypothesis $\mathrm{H}_1$ is true if $|\tilde{Y}_m|^2 \geq \tilde{\eta}$, otherwise the hypothesis $\mathrm{H}_0$ is claimed.

For the optimum DDL detector (A15), the probability of detection in the case of the Swerling I target is

$$P_{\mathrm{D}} = P_{\mathrm{FA}}^{\frac{1}{1+\tilde{\gamma}_{\mathrm{o}}}} \qquad \text{(A16)}$$

where $\tilde{\gamma}_{\mathrm{o}}$ is given by (A13).

**Appendix B**

Evaluation of the Probability of False Alarm

B1. GLR Detector

B1.1 Conventional Time Domain (TD) GLR Detector

The TD-GLR Detector is given by [1, 2]

$$\frac{1}{1+\frac{\mathbf{x}^{\mathrm{H}}\widehat{\mathbf{\Sigma}}^{-1}\mathbf{x}}{K_{\mathrm{T}}}}\frac{\left|\mathbf{s}^{\mathrm{H}}\widehat{\mathbf{\Sigma}}^{-1}\mathbf{x}\right|^2}{\mathbf{s}^{\mathrm{H}}\widehat{\mathbf{\Sigma}}^{-1}\mathbf{s}} \underset{\mathrm{H}_0}{\overset{\mathrm{H}_1}{\gtrless}} \lambda_{\mathrm{GLR}} = K_{\mathrm{T}}\xi_{\mathrm{T}} \qquad \text{(B1)}$$

where $\mathbf{s}$ is the $N \times 1$ known target steering vector (since the target Doppler frequency is known), and $\mathbf{x}$ is the $N \times 1$ vector associated with the cell under test (CUT). In this equation, $\widehat{\mathbf{\Sigma}}$ is the estimate of the interference covariance matrix $\mathbf{\Sigma}$ (unknown). This estimate is the sample covariance matrix based on a set of the independent training vectors $\mathbf{x}_k$, $k = 1,2,\dots,K_{\mathrm{T}}$, $K_{\mathrm{T}} > N$.

It is assumed that under hypothesis $\mathrm{H}_0$ (target is not present), the CUT vector and the training vectors share the same $N \times N$ covariance matrix $\mathbf{\Sigma}$, and the superscript $^{\mathrm{H}}$ stands for the Hermitian transposition.

The probability of false alarm for the TD-GLR detector in (B1) is given by [2]

$$P_{\mathrm{FA}} = \frac{1}{(1+\alpha_{\mathrm{T}})^L} \qquad \text{(B2)}$$

where $L = K_{\mathrm{T}} - N + 1$ and $\alpha_{\mathrm{T}} = \xi_{\mathrm{T}}/(1-\xi_{\mathrm{T}})$ with $\xi_{\mathrm{T}}$ being the factor for the threshold in (B1).

B1.2. Doppler Domain Localized (DDL) GLR Detector

The DDL-GLR detector of the order $n$ associated with the *m*-th RPTD is given by (15) as

$$\frac{1}{1+\frac{\tilde{\mathbf{y}}_m^{\mathrm{H}}\widehat{\tilde{\mathbf{\Phi}}}_m^{-1}\tilde{\mathbf{y}}_m}{K}}\frac{\left|\hat{\tilde{\mathbf{t}}}_m^{\mathrm{H}}\widehat{\tilde{\mathbf{\Phi}}}_m^{-1}\tilde{\mathbf{y}}_m\right|^2}{\hat{\tilde{\mathbf{t}}}_m^{\mathrm{H}}\widehat{\tilde{\mathbf{\Phi}}}_m^{-1}\hat{\tilde{\mathbf{t}}}_m} \underset{\mathrm{H}_0}{\overset{\mathrm{H}_1}{\gtrless}} \tilde{\lambda}_{\mathrm{GLR}} = K\tilde{\xi} \qquad \text{(B3)}$$

where $\widehat{\tilde{\mathbf{\Phi}}}_m$ is the sample DDL covariance matrix, see (14).

The DFT transform is a linear invertible transform that preserves Gaussianity with the only change of the received data vector, the signal steering vector and the disturbance covariance matrix. Therefore, the probability of false alarm equation for the DDL-GLR detector (B3) is identical to that for the TD-GLR detector. The only modification one must apply is the corresponding substitutions for the parameters.

Thus, the probability of false alarm for the DDL-GRL detector is given by

$$P_{\mathrm{FA}} = \frac{1}{(1+\tilde{\alpha})^{Q}} \tag{B4}$$

where $Q = K - n + 1$, and $\tilde{\alpha} = \tilde{\xi}/(1 - \tilde{\xi})$ with $\tilde{\xi}$ being the factor for the threshold in (B3).

B2. AMF Detector

B2.1 Conventional Time Domain AMF Detector

The TD-AMF Detector is given by [2]

$$\frac{\left|\mathbf{s}^{\mathrm{H}}\widehat{\mathbf{\Sigma}}^{-1}\mathbf{y}\right|^{2}}{\mathbf{s}^{\mathrm{H}}\,\widehat{\mathbf{\Sigma}}^{-1}\mathbf{s}} \underset{\mathrm{H}_0}{\overset{\mathrm{H}_1}{\gtrless}} \lambda_{\mathrm{AMF}} = K_{\mathrm{T}}\alpha_{\mathrm{T}} \tag{B5}$$

where the symbols $\mathbf{s}$, $\mathbf{y}$, and $\widehat{\mathbf{\Sigma}}$ are defined above in B1.1.

The probability of false alarm for the TD-AMF detector in (B5) is given by [2]

$$P_{\mathrm{FA}} = \int_0^1 \frac{f_{\beta}(\rho;\, L+1, N-1)}{(1+\alpha_{\mathrm{T}})^{L}}\, d\rho \tag{B6}$$

where $L = K_{\mathrm{T}} - N + 1$, $\alpha_{\mathrm{T}}$ is the factor for the threshold in (B6), and the central Beta density function is

$$f_{\beta}(x; n, m) = \frac{(n+m-1)!}{(n-1)!(m-1)!} x^{n-1}(1-x)^{m-1} \tag{B7}$$

For a given $P_{\mathrm{FA}}$, one can find the scalar $\alpha_{\mathrm{T}}$ from (B6) using numerical iterations.

B2.2. Doppler Domain Localized (DDL) AMF Detector

The DDL-AMF detector of the order $n$ associated with the *m*-th RPTD is given by (16) as

$$\frac{\left|\hat{\mathbf{t}}_{m}^{\mathrm{H}}\,\widehat{\mathbf{\Phi}}_{m}^{-1}\tilde{\mathbf{y}}_{m}\right|^{2}}{\hat{\mathbf{t}}_{m}^{\mathrm{H}}\,\widehat{\mathbf{\Phi}}_{m}^{-1}\hat{\mathbf{t}}_{m}} \underset{\mathrm{H}_0}{\overset{\mathrm{H}_1}{\gtrless}} \tilde{\lambda}_{\mathrm{AMF}} = K\tilde{\alpha} \tag{B8}$$

Since the DFT preserves Gaussianity, the equation for the probability of false alarm for the DDL-AMF detector (B8) is identical to that for the detector (B5). Hence, the corresponding substitutions for the parameters yield

$$P_{\mathrm{FA}} = \int_0^1 \frac{f_\beta(\rho;\, Q+1, n-1)}{(1+\tilde{\alpha})^Q} d\rho \tag{B9}$$

where $Q = K - n + 1$, and $\tilde{\alpha}$ is the factor for the threshold in (B8). One should use numerical iterations to find the scalar $\tilde{\alpha}$ from (B9) for a given $P_{\mathrm{FA}}$.

**Appendix C**

Evaluation of the Probability of Detection

This Appendix assumes the Swerling I target model and known target Doppler frequency.

C1. GLR Detector

C1.1 Conventional Time Domain (TD) GLR Detector

For the conventional TD-GLR Detector given by (B1), the probability of detection $P_{\mathrm{D}}$ is provided by the following integral expression [2]

$$P_{\mathrm{D}} = \int_0^1 \left(\frac{1+\gamma_{\mathrm{o}}\rho}{1+\alpha_{\mathrm{T}}+\gamma_{\mathrm{o}}\rho}\right)^L f_\beta(\rho; L+1, N-1) d\rho \tag{C1}$$

where $\gamma_{\mathrm{o}} = \gamma_{\mathrm{in}}\mathbf{s}^{\mathrm{H}}\mathbf{\Sigma}^{-1}\mathbf{s}$ is the optimum filter output SDR, with $\gamma_{\mathrm{in}}$ being the SDR at the input (see Subsection A1). Other parameters are: $L = K_{\mathrm{T}} - N + 1$, the scalar $\alpha_{\mathrm{T}}$ is determined from (B2) for the specified probability of false alarm $P_{\mathrm{FA}}$ as $\alpha_{\mathrm{T}} = P_{\mathrm{FA}}^{-1/L} - 1$, and $f_\beta(x; n, m)$ is the central Beta density function given by (B7).

C1.2 Doppler Domain Localized (DDL) GLR Detector

Since the DFT preserves Gaussianity, the probability of detection for the DDL-GLR detector given by (B3) follows from (C1) after proper symbol substitutions as

$$P_{\mathrm{D}} = \int_0^1 \left(\frac{1+\tilde{\gamma}_{\mathrm{o}}\rho}{1+\tilde{\alpha}+\tilde{\gamma}_{\mathrm{o}}\rho}\right)^Q f_\beta(\rho; Q+1, n-1) d\rho \tag{C2}$$

where $\tilde{\gamma}_{\mathrm{o}} = \gamma_{\mathrm{in}}(\tilde{\mathbf{t}}_m^{\mathrm{H}}\tilde{\mathbf{\Phi}}_{0m}^{-1}\tilde{\mathbf{t}}_m)$ is the optimum DDL filter output SDR, with $\gamma_{\mathrm{in}}$ being the SDR at the input (see Subsection A2). Other parameters are: $Q = K - n + 1$, the scalar $\tilde{\alpha}$ is determined from (B4) for the specified probability of false alarm $P_{\mathrm{FA}}$ as $\tilde{\alpha} = P_{\mathrm{FA}}^{-1/Q} - 1$.

C2. AMF Detector

C2.1 Conventional Time Domain (TD) AMF Detector

For the conventional TD-AMF Detector given by (B5), the probability of detection $P_{\mathrm{D}}$ is provided by the following integral expression [2]

$$P_{\mathrm{D}} = \int_0^1 \left(\frac{1+\gamma_{\mathrm{o}}\rho}{1+(\alpha_{\mathrm{T}}+\gamma_{\mathrm{o}})\rho}\right)^L f_\beta(\rho; L+1, N-1)d\rho \tag{C3}$$

where $\gamma_{\mathrm{o}} = \gamma_{\mathrm{in}}\mathbf{s}^{\mathrm{H}}\mathbf{\Sigma}^{-1}\mathbf{s}$ is the optimum filter output SDR, with $\gamma_{\mathrm{in}}$ being the SDR at the input (see Subsection A1). Other parameters are: $L = K_{\mathrm{T}} - N + 1$ and the scalar $\alpha_{\mathrm{T}}$ is determined from (B6) for the specified probability of false alarm $P_{\mathrm{FA}}$ using numerical iterations.

C2.2 Doppler Domain Localized (DDL) AMF Detector

Since the DFT transform preserves Gaussianity, the probability of detection for the DDL-AMF detector given by (B8) follows from (C3) after proper symbol substitutions as

$$P_{\mathrm{D}} = \int_0^1 \left(\frac{1+\tilde{\gamma}_{\mathrm{o}}\rho}{1+(\tilde{\alpha}+\tilde{\gamma}_{\mathrm{o}})\rho}\right)^Q f_\beta(\rho; Q+1, n-1)d\rho \tag{C4}$$

where $\tilde{\gamma}_{\mathrm{o}} = \gamma_{\mathrm{in}}(\tilde{\mathbf{t}}_m^{\mathrm{H}}\widetilde{\mathbf{\Phi}}_{0m}^{-1}\tilde{\mathbf{t}}_m)$ is the optimum DDL filter output SDR, with $\gamma_{\mathrm{in}}$ being the SDR at the input (see Subsection A2). Other parameters: $Q = K - n + 1$, the scalar $\tilde{\alpha}$ is determined from (B9) for the specified probability of false alarm $P_{\mathrm{FA}}$ using numerical iterations.

## Appendix D

DDL-AMF Detector: Simple and Accurate Threshold Calculation

For the DDL-AMF detector, calculating the detection threshold $\tilde{\lambda}$ from (B9) entails time-consuming numerical iterations. This approach is not acceptable for a real-time implementation of DDL-AMF detectors. Appendix C provides an efficient and accurate approximating formula for computing $\tilde{\lambda}$.

The approximating equation for a given $P_{\mathrm{FA}}$ is given by

$$\tilde{\lambda}_a(\mathbf{c}) = c_1 P_{FA}^{(-1/c_2)} - c_3 \tag{D1}$$

where $\tilde{\lambda}_a(\mathbf{c})$ is the approximate DDL-AMF detector's threshold and $\mathbf{c} = [c_1, c_2, c_3]$ is the vector of approximating coefficients to be sought. We define the optimum vector $\mathbf{c} = \mathbf{c}_{\mathrm{opt}}$ as the solution to the following minimax problem

$$\mathbf{c}_{\text{opt}} = \arg\min_{\mathbf{c}} \max\left(\left|\tilde{\boldsymbol{\Lambda}} - \tilde{\boldsymbol{\Lambda}}_a(\mathbf{c})\right| ./\, \tilde{\boldsymbol{\Lambda}}\right)$$
$$\mathbf{c}_{\text{o}} = [1, K, 1] \tag{D2}$$

where the symbols "./" represent the elementwise division, and $\mathbf{c}_{\text{o}}$ is the initial approximation, with $K$ being the number of training vectors.

In (D2), the vector $\tilde{\boldsymbol{\Lambda}} = [\tilde{\lambda}_1\ \tilde{\lambda}_2 \ldots \tilde{\lambda}_Q]$ is the vector of exact DDL-AMF thresholds precomputed using (B9) for the vector of the reference $P_{\text{FA}}$ values $\mathbf{P} = [P_{\text{FA}1}\ P_{\text{FA}2} \ldots P_{\text{FA}Q}]$; $Q$ is the length of $\mathbf{P}$. The vector $\tilde{\boldsymbol{\Lambda}}_a(\mathbf{c}) = [\tilde{\lambda}_{a1}(\mathbf{c})\ \tilde{\lambda}_{a2}(\mathbf{c}) \ldots \tilde{\lambda}_{aQ}(\mathbf{c})]$ represents the approximate DDL-AMF thresholds, where the $q$-th value $\tilde{\lambda}_{aq}(\mathbf{c})$, $q = 1,2,\ldots,P$, is computed from (D1) for the corresponding $P_{\text{FA}q}$ values in $\mathbf{P}$.

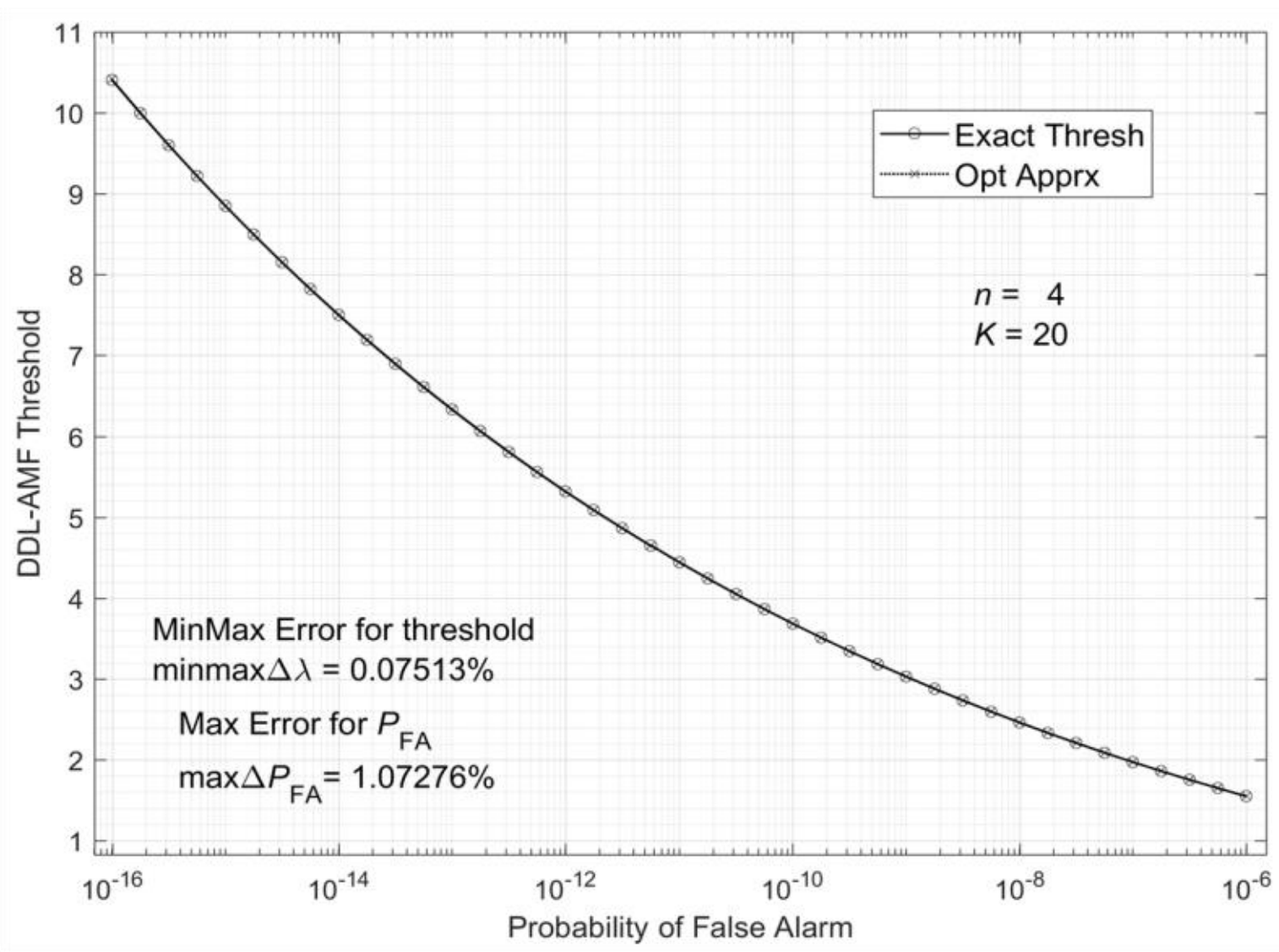

Fig. D1. Comparison of exact and approximate DDL-AMF thresholds

Fig. D1 compares the exact DDL-AMF thresholds $\tilde{\boldsymbol{\Lambda}}$ and the optimum approximate DDL-AMF thresholds $\tilde{\boldsymbol{\Lambda}}_a(\mathbf{c}_{\text{opt}})$ computed as functions of the probability of false alarm for $n = 4$, $K = 20$ for the reference vector $\mathbf{P}$ of size $1 \times 41$ defined over the interval [$10^{-16}$, $10^{-6}$] as

**P** = unique([logspace(-16, -15, 5), logspace(-15, -14, 5), logspace(-14, -13, 5), logspace(-13, -12, 5), logspace(-12, -11, 5), logspace(-11, -10, 5), logspace(-10, -9, 5), logspace(-9, -8, 5), logspace(-8, -7, 5), logspace(-7, -6, 5)]) (D3) where "unique" and "logspace" are Matlab functions.

Under these settings, the optimum vector $\mathbf{c}_{\text{opt}}$ is found to be (in double precision format)

$$\mathbf{c}_{\text{opt}} = [1.137593213858974,\ \ 15.875828450315428,\ \ 1.168722472188503]$$

Fig. D1 shows that the exact and optimum DDL-AMF thresholds are in excellent agreement (their plots are not visually distinguished) because the relative error $\Delta\tilde{\lambda}$ between them does not exceed the minimax error $100 \times \max\left(\left|\tilde{\boldsymbol{\Lambda}} - \tilde{\boldsymbol{\Lambda}}_a(\mathbf{c}_{\text{opt}})\right| ./\, \tilde{\boldsymbol{\Lambda}}\right) = 0.07513\%$. The relative error $\Delta P_{\text{FA}}$ between the exact $P_{\text{FA}}$ values and approximate $P_{\text{FA}}$ values computed from (D1) using approximate thresholds from the vector $\tilde{\boldsymbol{\Lambda}}_a(\mathbf{c}_{\text{opt}})$ is also small – it does not exceed 1.07276%.

For $n = 4$ and 5, Table DI summarizes the optimum approximating coefficients for computing corresponding DDL-AMF thresholds. The coefficients are calculated for $K = 3n, 4n$, and $5n$ using the reference vector in (D3).

TABLE DI
OPTIMUM APPROXIMATING COEFFICIENTS FOR COMPUTING DDL-AMF THRESHOLDS
REFERENCE $P_{\text{FA}}$ VALUES ARE IN THE INTERVAL [$10^{-16}$, $10^{-6}$] AS SPECIFIED BY (D3)

| $n$ | $K$ | MinMax $\Delta\tilde{\lambda}$ (%) | Max $\Delta P_{\text{FA}}$(%) | $c_{1\text{opt}}$ | $c_{2\text{opt}}$ | $c_{3\text{opt}}$ |
|---|---|---|---|---|---|---|
| 4 | 12 | 0.34064 | 2.90489 | 1.514702363388335 | 8.702739376483137 | 1.915367688856702 |
| | 16 | 0.14638 | 1.69826 | 1.245454831927344 | 12.252243482999409 | 1.337256680537876 |
| | 20 | 0.07513 | 1.07276 | 1.137593213858974 | 15.875828450315428 | 1.168722472188503 |
| 5 | 15 | 0.27319 | 2.67988 | 1.479403713377666 | 10.424987325960629 | 1.726947663133344 |
| | 20 | 0.10083 | 1.29530 | 1.224205523025362 | 14.809425915487832 | 1.280983519608858 |
| | 25 | 0.04777 | 0.78369 | 1.122972523283579 | 19.307882309923791 | 1.141253119783500 |

**Appendix E**

"Rapid Adaptive Matched Filter for Detecting Radar Targets With Unknown Velocity,"
in *IEEE Access*, vol. 12, pp. 25411-25428, 2024, doi: 10.1109/ACCESS.2024.3363242.

**Errata**

Page 25415:

Equation (9) should be $\eta_{lm} = \frac{1}{\tilde{\mathbf{s}}_{lm}^{\text{H}}\hat{\boldsymbol{\Sigma}}_l^{-1}\mathbf{s}_{lm}} \frac{|\tilde{\mathbf{s}}_{lm}^{\text{H}}\hat{\boldsymbol{\Sigma}}_l^{-1}\tilde{\mathbf{y}}_l|^2}{1+(\tilde{\mathbf{y}}_l^{\text{H}}\hat{\boldsymbol{\Sigma}}_l^{-1}\tilde{\mathbf{y}}_l)/K}$

In the 2nd column, 2nd paragraph, 4th and 5th lines, the text fragment should be frequency) for the 4th order RODI-based DDL-GLR detector, see [4, p. 536, Fig. 3]. Analyzing the frequency response

In the 2nd column, last paragraph, 4th line, the text fragment should be

In this figure, the number of pulses and training samples

Page 25420:

Equation (15) should be $\widetilde{\Lambda}_{\mathrm{GLR}m} = \frac{1}{1+(\tilde{\mathbf{y}}_m^{\mathrm{H}}\widehat{\boldsymbol{\Sigma}}_m^{-1}\tilde{\mathbf{y}}_m)/K}\frac{\left|\hat{\mathbf{t}}_m^{\mathrm{H}}\,\widehat{\boldsymbol{\Phi}}_m^{-1}\tilde{\mathbf{y}}_m\right|^2}{\hat{\mathbf{t}}_m^{\mathrm{H}}\,\widehat{\boldsymbol{\Phi}}_m^{-1}\hat{\mathbf{t}}_m} \underset{\mathrm{H}_0}{\overset{\mathrm{H}_1}{\gtrless}} \tilde{\lambda}_{\mathrm{GLR}}$

Equation (16) should be $\widetilde{\Lambda}_{\mathrm{AMF}m} = \frac{\left|\hat{\mathbf{t}}_m^{\mathrm{H}}\,\widehat{\boldsymbol{\Phi}}_m^{-1}\tilde{\mathbf{y}}_m\right|^2}{\hat{\mathbf{t}}_m^{\mathrm{H}}\,\widehat{\boldsymbol{\Phi}}_m^{-1}\hat{\mathbf{t}}_m} \underset{\mathrm{H}_0}{\overset{\mathrm{H}_1}{\gtrless}} \tilde{\lambda}_{\mathrm{AMF}}$

Page 25424:

Equation (A7) should be $|Y|^2 \underset{\mathrm{H}_0}{\overset{\mathrm{H}_1}{\gtrless}} \eta$

Page 25425:

Equation (A15) should be $|\tilde{Y}_m|^2 \underset{\mathrm{H}_0}{\overset{\mathrm{H}_1}{\gtrless}} \tilde{\eta}$

Equation (B1) should be $\frac{1}{1+\frac{\mathbf{x}^{\mathrm{H}}\,\widehat{\boldsymbol{\Sigma}}^{-1}\mathbf{x}}{K_{\mathrm{T}}}}\frac{|\mathbf{s}^{\mathrm{H}}\widehat{\boldsymbol{\Sigma}}^{-1}\mathbf{x}|^2}{\mathbf{s}^{\mathrm{H}}\,\widehat{\boldsymbol{\Sigma}}^{-1}\mathbf{s}} \underset{\mathrm{H}_0}{\overset{\mathrm{H}_1}{\gtrless}} \lambda_{\mathrm{GLR}} = K_{\mathrm{T}}\xi_{\mathrm{T}}$

Page 25426:

Equation (B3) should be $\frac{1}{1+\frac{\tilde{\mathbf{y}}_m^{\mathrm{H}}\widehat{\boldsymbol{\Phi}}_m^{-1}\tilde{\mathbf{y}}_m}{K}}\frac{\left|\hat{\mathbf{t}}_m^{\mathrm{H}}\,\widehat{\boldsymbol{\Phi}}_m^{-1}\tilde{\mathbf{y}}_m\right|^2}{\hat{\mathbf{t}}_m^{\mathrm{H}}\,\widehat{\boldsymbol{\Phi}}_m^{-1}\hat{\mathbf{t}}_m} \underset{\mathrm{H}_0}{\overset{\mathrm{H}_1}{\gtrless}} \tilde{\lambda}_{\mathrm{GLR}} = K\tilde{\xi}$

Equation (B5) should be $\frac{|\mathbf{s}^{\mathrm{H}}\widehat{\boldsymbol{\Sigma}}^{-1}\mathbf{y}|^2}{\mathbf{s}^{\mathrm{H}}\,\widehat{\boldsymbol{\Sigma}}^{-1}\mathbf{s}} \underset{\mathrm{H}_0}{\overset{\mathrm{H}_1}{\gtrless}} \lambda_{\mathrm{AMF}} = K_{\mathrm{T}}\alpha_{\mathrm{T}}$

Equation (B8) should be $\frac{\left|\hat{\mathbf{t}}_m^{\mathrm{H}}\,\widehat{\boldsymbol{\Phi}}_m^{-1}\tilde{\mathbf{y}}_m\right|^2}{\hat{\mathbf{t}}_m^{\mathrm{H}}\,\widehat{\boldsymbol{\Phi}}_m^{-1}\hat{\mathbf{t}}_m} \underset{\mathrm{H}_0}{\overset{\mathrm{H}_1}{\gtrless}} \tilde{\lambda}_{\mathrm{AMF}} = K\tilde{\alpha}$

The right-hand part in equation (B5) should be $K_{\mathrm{T}}\alpha_{\mathrm{T}}$

The text fragment under equation (B6) should be
where $L = K_{\mathrm{T}} - N + 1$, $\alpha_{\mathrm{T}}$ is the factor for the threshold in (B6), and the central Beta density function is

The right-hand part in equation (B8) should be $K\tilde{\alpha}$

The text fragment under equation (B9) should be
where $Q = K - n + 1$, and $\tilde{\alpha}$ is the factor for the threshold in (B8). One should use numerical iterations to find the scalar $\tilde{\alpha}$ from (B9) for a given $P_{\mathrm{FA}}$.

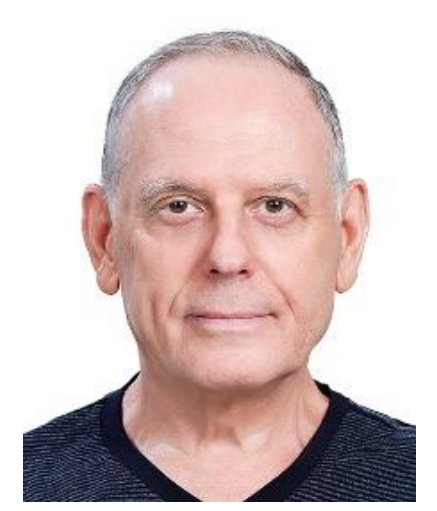

**Anatolii A. Kononov** (Independent Researcher) received the M. Sc. (with honors) degree in electrical engineering from the Odesa National Polytechnic University, Odesa, Ukraine, in 1975, and the Ph.D. degree in electrical engineering from the Electrotechnical University “LETI,” Saint Petersburg, Russia, in 1983. From 1975 to 2002, he was at the Odesa National Polytechnic University, first as a Research Engineer and later as an Associate Professor with the Department of Radio Engineering. From 2002 to 2005 and 2010 to 2022, he was a Senior Researcher with the Research Center at STX Engine, Yongin, Korea. From 2006 to 2008, he was with the Tech University of Korea (TU Korea), Siheung, Korea, as an Invited Professor with the Department of Electronic Engineering. From 2008 to 2010, he was a Senior Researcher with the Department of Earth and Space Sciences at the Chalmers University of Technology, Gothenburg, Sweden. Since 2023 he has been with the School of Integrated Technology, Yonsei University, Seoul, Korea, as a Senior Researcher. His main research interests are in radar signal processing, system analysis and design.